\documentstyle [12pt]{article}
\begin{document}

\title{Evolving Maximally Symmetric Spacetime Bubbles from
Spontaneous $Z_2-$Violation at Electroweak Symmetry Breaking
Scale}

\vspace{1cm}

\author{Ciprian Dariescu\footnote{on leave of absence from
Dept. of Theoretical Physics, ^^ ^^ Al. I. Cuza'' University,
Ia\c{s}i, 6600, Romania}
\\ {\it Institute of Theoretical Science}
\\ {\it 5203 University of Oregon}
\\ {\it Eugene OR 97403-5203}
\\
email: ciprian@physics.uoregon.edu}

\date{\today}
\maketitle

\begin{abstract}
The Einstein-Gordon equations for Friedmann-Robertson-Walker (FRW)
geometries in feedback reaction with the quartically
self-in\-te\-rac\-ting physical field, arisen from the ^^ ^^ inner
parity'' spontaneous breaking, are explicitly formulated. The
Hamiltonian density non-positive extrema would classically forbid
both spatially closed and flat homogeneous and isotropic worlds if
these were to allow the physical field to (repeatedly) go through
and to (finally) settle down in a ground state. In this respect,
the {\it fixed point} exact solutions of the spontaneous
$Z_2-$symmetry breaking Einstein-Gordon equations (mandatory)
describe $(k=-1)-$FRW manifolds which actually are either Milne or
anti-de Sitter Universes. Setting the $Z_2-$invariance breaking
scale at the one of the electroweak symmetry, we speculate on the
cosmological implications of the Higgs-anti-de Sitter bubbles and
derive a set of particular closed-form solutions to the
$S^2-$cobordism with a spatially-flat FRW Universe.
\end{abstract}

\vspace*{0.5cm}

{\it Keywords}:

\begin{itemize}
\item{} Friedmann-Robertson-Walker Cosmology, \item{}
3-dimensional spacelike manifold of negative curvature, \item{}
spontaneously broken $Z_{2}-$symmetry, \item{} ^^ ^^ field
reflection'' non-invariant Einstein-Gordon equations, \item{}
Milne and anti-de Sitter spacetimes, \item{} electroweak symmetry
breaking scale, Higgs boson, \item{} CMBR, galaxies and quasars,
Inflation, \item{} Higgs$-$anti-de Sitter bubbles, \item{}
$S^{2}-$cobordism.
\end{itemize}

PACS numbers: 11.27.+d, 04.40.Nr., 04.20.Jb
\\
\baselineskip 1.5em

\newpage

\section{Introduction}

The resourceful M Theory [1-10] and the celebrated AdS/CFT
correspondence [11,12] have released a high tide of (new)
investigations on the anti-de Sitter (AdS) spacetime, its extended
versions to more than four dimensions and on the features of the
(quantum) matter-fields evolving therein [13-22]. The applications
to High Energy Physics, General Relativity and Cosmology
[14,23-32], covering topics like 4D gauge field theories {\it via}
M Theory, ^^ ^^ conformal Higgs'', mass hierarchy (problem),
dimensional compactification, either small or large extra
dimensions, new unified theories, etc., together with the late
observational data on the Universe (cosmological) acceleration
[33-37], determinedly go far beyond the (so-called) Standard
Model(s). Along with the growing interest in Pre-Big-Bang and
other new types of Inflation [38-49] $-$which could also solve the
{\it cosmological constant problem}$-$ and in the models with
various ^^ ^^ dark energy'' contents [48,50], there is also a need
for a better understanding of the geometrodynamical link between
the spacetime structure and the nature of spontaneously symmetry
breaking vacua. A comprehensive account on this matter,
emphasizing the False Vacuum Physics ^^ ^^ subtleties'' and
quoting most of the previous papers in the field, has been given
by T. Banks [51]; somewhat similarly, for an external continuous
symmetry, J. W. Moffat [52] has explicitly worked out the
intriguing, particularly astrophysical, consequences of the
$SO(3,1)-$invariance spontaneous breaking. Last but not least,
with respect to the subject we would like to speculate on, there
is also the recent paper of A. Dev {\it et al.} [53], where the
conclusion is drawn, based on a careful analysis of the late(st)
{\it gravity lensing} and {\it high-z supernova} data [54], that
the possibility of a (quasi)Milne stage, i.e. $(k=-1)-$FRW
spacetime with linearly expanding scale function (so that the
Universe could undergo a uniform expansion), has actually not been
completely ruled out.

Formally, what we are dealing with, in the present paper, is a
geometrodynamical analysis of the ^^ ^^ extremal'' spacetime
structures derived as exact solutions of the Einstein's field
equations, for (initially) a generic FRW background with a
quartically self-interacting scalar field as matter-content, in
the system fixed points, i.e. at the three $Z_2-$symmetric extrema
of the (fourth-degree polynomial) Hamiltonian density.
The ^^ ^^ catch'' is that, while the central fixed point, the local
maximum, is ^^ ^^ gravitationally'' inconsistent {\it only} with
the {\it spatially-closed} FRW geometries, {\it each} of the {\it
other} two {\it fixed points}, the absolute minima representing
the matter-source degenerate vacua, is (geometrodynamically)
consistent (in the sense of an $R-$valued solution to the
corresponding Einstein equations) {\it only} with the
$(k=-1)-${\it family} of FRW manifolds. As a matter of fact, once
the vacuum has been set in one of the (two possible) ground
states, the spontaneous$Z_2-$symmetry breaking does
^^ ^^ instantly'' create an anti-de Sitter Universe; when slightly
perturbed, it gets filled with massive particles representing the
physical field (quantum) excitations around the settled ground
state. Considering therefore ^^ ^^ $k=-1$'' as a {\it compulsory}
condition, the previously mentioned {\it central fixed point}
corresponds to a {\it Milne phase}, which, being unstable against
the coherent field fluctuations, does primarily turn into an {\it
anti-de Sitter one} of (very) small curvature (in absolute value).

Informally from the rigour perspective, we set the
$Z_2-$invariance breaking scale at the one of the electroweak
symmetry,^^ ^^ taking'' the Higgs-boson mass somewhere inbetween
$115$ and $300 \, GeV$, and analyzing the respective values for
the ^^ ^^ gravitationally sustained'' proper-pulsation, energy and
power of the Higgs$-$anti-de Sitter bubbles, we speculate on some
of their cosmological implications, such as a stronger CBR
anisotropy on the frequencies ranging from (about) $190 \, MHz$ upto
(some) $1.4 \, GHz$, more prominent towards the Giant Void(s),
seizable deviations from the ^^ ^^ whole sky''-averaged
intensity-level of the $21 \, cm(s)$ Hydrogen line, inner parity
violating seeds of galaxy formation and Higgs-vacuum-based anti-de
Sitter power-sources feeding the quasars' cores.

A particular set of closed-form solutions to the $S^2-$cobordism
between an anti-de Sitter bubble and a spatially-flat Universe is
readily worked out in the final part of the paper. It
(generically) points out that, as seen {\it from} the $(k=0)-$FRW
spacetime, the coordinate-radius of the small anti-de Sitter
bubbles, as well as of the ones (not necessarily small) that might
have existed in the course of Inflation, does asymptotically
vanish at some high exponential rate. However, analytically,
similar conclusions on the {\it large} bubbles evolution, in a
{\it subexponentially} expanding conformally-flat Universe, cannot
be drawn so easily, because of the highly nonlinear character of
the respective $S^2-$cobordering equation(s).

\section{Spontaneously Broken ${\mbox{\boldmath $Z_2-$}}$Symmetry}

Let us consider the inner parity (i.e. the field reflection $\Phi
\rightarrow - \, \Phi$) invariant Lagrangian density
\begin{equation}
{\cal L}[ \Phi ]\, = \, - \, \frac{1}{2} \, \eta^{ab} \, \Phi_{|a}
\Phi_{|b} \, + \, \frac{1}{2} \, \mu^2 \, \Phi^2 \, - \,
\frac{\lambda}{24} \, \Phi^4
\end{equation}
of a quartically self-interacting real scalar field $\Phi$, where,
${\mbox{\boldmath $\eta$}}\, = \, diag[1,1,1,-1]$ is the
fundamental metric tensor for a pseudo-orthonormal tetrad $\lbrace
e_a = e_{a}^{i} \, \partial_i \rbrace_{a=\overline{1,4}}^{i=
\overline{1,4}}$ whose dual $\lbrace \omega^a = \omega^{a}_{i} \,
dx^i \,\, | \,\, \langle \omega^a \, , \, e_b \rangle =
\delta^{a}_{b} \, \rbrace^{i=\overline{1,4}}_{a,b=\overline{1,4}}$
generates the spacetime metric
\begin{equation}
ds^2 \, = \, \eta_{ab} \, \omega^a \, \omega^b \,\, ,
\end{equation}
the $( \, \cdot \, )_{|a}$ notation stands for the tetradic
derivative with respect to $e_a$, i.e. $( \, \cdot \, )_{|a} \, =
\, e_{a} ( \, \cdot \, ) \, =\, e_{a}^{i} \, \partial_{i}( \,
\cdot \, )$, and $\mu^2$ $-$ with ${\rm mass}^2$ dimension $-$ and
$\lambda$ (dimensionless) are the two positive parameters that
^^ ^^ accommodate'' the spontaneously (discrete) symmetry breaking
mechanism. Working out, from the functional expression
\begin{equation}
{\mbox{\boldmath $T$}} \, = \, - \, \frac{2}{\sqrt{- \, g}} \,
\frac{\delta}{\delta {\mbox{\boldmath $g$}}} \left[ \sqrt{- \, g}
\, {\cal L} \right] \, \, ,
\end{equation}
the covariant components (with respect to the ^^ ^^ rigid'' tensor
basis \\
$\lbrace \omega^a \otimes \omega^b
\rbrace_{a,b=\overline{1,4}}\,$) of the conservative
stress-energy-momentum tensor, for the considered scalar field, it
results in full
\begin{equation}
T_{ab} \, = \, \Phi_{|a} \, \Phi_{|b} \, - \, \frac{1}{2} \,
\eta_{ab} \, \left[ \Phi^{|c} \, \Phi_{|c} \, - \, \mu^2 \, \Phi^2
\, + \, \frac{\lambda}{12} \, \Phi^4 \, \right] \, \, ,
\end{equation}
so that, the extrema of the Hamiltonian density
\begin{equation}
{\cal H} \, = \, T_{44} \, = \, \frac{1}{2} \, \delta^{ab} \,
\Phi_{|a} \, \Phi_{|b} \, - \, \frac{\mu^2}{2} \, \Phi^2 \, + \,
\frac{\lambda}{24} \, \Phi^4
\end{equation}
are given by the equation
\begin{equation}
\frac{\partial {\cal H}}{\partial \Phi} \left(
\, \Phi_0 \, \right) \, = \,
\Phi_0 \,\left( \frac{\lambda}{6} \, \Phi_0^2 \,
- \, \mu^2 \right) \, = \,0 \,\, ,
\end{equation}
i.e. $\lbrace \Phi_{0}^{\alpha} \rbrace_{\alpha=-,0,+} = \left
\lbrace - \mu \sqrt{\frac{6}{\lambda}}, \, 0 \, , \, \mu
\sqrt{\frac{6}{\lambda}} \right \rbrace$. Inspecting the sign of
the second derivative
\begin{equation}
\frac{\partial^{2}{\cal H}}{\partial \Phi^2} \left( \, \Phi_0 \,
\right)\, = \,
\frac{\lambda}{2}\, \Phi_0^2 \, - \, \mu^2
\end{equation}
for each of the three roots $\lbrace \Phi_{0}^{\alpha}
\rbrace_{\alpha=-,0,+}$, it instantly results that $\Phi_{0}^{0}$,
where $\frac{\partial^{2}{\cal H}}{\partial \Phi^2}= \,- \mu^2$,
is an {\it unstable} fixed point, while $\Phi_{0}^{\pm}$, where
$\frac{\partial^{2} {\cal H}}{\partial \Phi^2}= \,2 \mu^2$, are
the ^^ ^^ real'' {\it minima} which correspond to the two possible
ground states of the initially fictitious (i.e. apparently
deprived of direct particle interpretation) scalar field $\Phi$.

Choosing $v = \Phi_{0}^{+} = \mu \sqrt{\frac{6}{\lambda}}$ as the
vacuum expectation value of $\Phi$, in its ground state, and
accordingly shifting the field
\begin{equation}
\Phi \, = \, v \, + \, \chi \, , \; \; {\rm where} \; \; \chi :
M_4 \rightarrow {\rm{\bf R}} \, ,
\end{equation}
such that $\chi = 0$ represents the {\it true} vacuum of the
theory, one ends up with the spontaneously ${\mbox{\boldmath
$Z$}}_2$ broken Lagrangian density
\begin{equation}
{\cal L}[ \chi ] \, = \, - \, \frac{1}{2} \, \chi^{|c} \, \chi_{|c} -
\,\frac{1}{2}(2 \mu^2) \chi^2 \, -\, \mu
\sqrt{\frac{\lambda}{6}}\, \chi^3 \, - \, \frac{\lambda}{24} \,
\chi^4 \,+ \, \frac{3 \, \mu^4}{2 \, \lambda}
\end{equation}
of the physically observable massive, $m_{\chi} \,= \, \sqrt{2} \,
\mu$, (real) scalar field $\chi$, subsequently obeying the inner
parity violating ternary nonlinear (generalized) Gordon equation
\begin{equation}
\Box \chi \, - \, \left( 2 \, \mu^2 \right) \, \chi \, = \,
3 \, \mu \sqrt{\frac{\lambda}{6}} \, \chi^2 \, + \,
\frac{\lambda}{6} \, \chi^3 \, \, ,
\end{equation}
where
\begin{equation}
\Box \chi \, = \, \frac{1}{\sqrt{- \, g}} \,
\frac{\partial}{\partial x^i} \left[\sqrt{- \, g} \,
g^{ik} \frac{\partial \chi}{\partial x^k} \right]
\end{equation}
is the d'Alembert operator on $M_4$, in terms of some local
coordinates $\lbrace x^i \rbrace_{i=\overline{1,4}}$.

Therefore, either from (9) and (3), or straightly from (4) with
the shift (8), it yields for the components of the energy-momentum
tensor ${\rm{\bf T}}$ (of the physical field $\chi$, and with
respect to the tensor basis $\lbrace \omega^a \otimes \omega^b
\rbrace_{a,b=\overline{1,4}}$) the actual expression
\begin{equation}
T_{ab} \, = \, \chi_{|a} \, \chi_{|b} \, - \,
\frac{1}{2}\, \eta_{ab} \, \left[ \chi^{|c} \, \chi_{|c} \, + \,
2 \, V(\chi) \, - \, \frac{3 \, \mu^4}{\lambda} \right] \, \, ,
\end{equation}
where the total, semi-classical (i.e. without quantum corrections)
potential
\begin{equation}
V(\chi) \, = \, \mu^2 \, \chi^2 \,+ \, \mu
\sqrt{\frac{\lambda}{6}} \, \chi^3 \, + \, \frac{\lambda}{24} \,
\chi^4
\end{equation}
is clearly no longer invariant under the discrete transformation
$\chi \, \rightarrow - \, \chi$.

In the Minkowski spacetime, which keeps on being flat whatever the
matter energy-momentum is, the time-translation isometry, and the
respective action-functional invariance, accounting for energy
conservation, allows gauging the energy scale by any constant
amount. Hence, any constant, field-independent, contribution to
the (44)-component of the conservative energy-momentum tensor,
such as $-3 \mu^4/(2 \lambda)$ in (12), does actually leave no
observable signature in the field dynamics and so, it can just
simply be thrown away. However, in general and physically more
realistic situations, where gravity cannot be neglected, the
matter stress-energy-momentum tensor does clearly affect the
metric of the Lorentzian base manifold, so that any of its
additional terms, even a constant one, cannot be omitted any
longer unless there are some serious, both mathematical and
physical, reasons.

\section{Inner Parity Non-Invariant Einstein-Gordon
Equations in FRW Cosmologies}

We have come to the point where we can address the question of
what type of cosmic-time evolving homogeneous and isotropic
3-geometry ^^ ^^ fits'' the massive scalar source-field $\chi$ in
such a way to produce an exact solution to the
Einstein-(generalized) Gordon equations. Since both homogeneity
and isotropy require a {\it maximal} $G_6-$group of motion on the
3-dimensional (sub)manifold $N_3$ (of $M_4$), this must possess
{\it constant} curvature, i.e. $k = \lbrace 1,0, \, -1 \rbrace$,
and thus, can only be the sphere $S^3$, the Euclidian
$\textbf{R}^3$, or the disconnected wings of the hyperboloid $H^3$
defined, in the flat $\textbf{R}^4$ of Cartesian coordinates
$\left( X^{\alpha}, \, T \right)_{\alpha=\overline{1,3}}$ and
metric signature 2, by the ^^ ^^ typical'' equation $T^2 \, - \,
\delta_{\alpha \beta}\, X^{\alpha} \, X^{\beta} \, = \, 1$. Hence,
in terms of dimensionless Euler-like coordinates
\begin{equation}
(\alpha, \beta, \theta) \, = \left \lbrace
\begin{array} {l}
(0,2 \, \pi) \times (0,2 \, \pi) \times (0, \frac{\pi}{2}
\,{\rm or} \, \pi) \; \; {\rm on} \; \; S^3 \, , k=1 \\
{\rm{\bf R}}^3 \, , k=0 \\
{\rm{\bf R}} \times (0,2 \, \pi) \times {\rm{\bf R}} \; \;
{\rm on} \; \; H^3
, \, k=-1 \\
\end{array} \right.
\end{equation}
the metric on $N_3$ does respectively read
\begin{equation}
 dl_{N_3}^2 \, = \left \lbrace
 \begin{array} {l}
 \cos^2 \theta \, (d \alpha)^2 \, + \, \sin^2 \theta \,
 (d \beta)^2 \, + \, (d \theta)^2 \, , \, k=1 \\
 (d \alpha)^2 \, + \, (d \beta)^2 \, + \, (d \theta)^2 \, , \,
 k=0 \\
 \cosh^2 \theta \, (d \alpha)^2 \, + \, \sinh^2 \theta \,
 (d \beta)^2 \, + \, (d \theta)^2 \, , k=-1 \\
\end{array} \right.
\end{equation}
and therefore, considering the spacetime $M_4 \, = \, N_3 \,
\times \, {\rm{\bf R}}$ as a continuous ^^ ^^ tower'' of $\lbrace
t = cst \, | \, \forall \, cst \in R \rbrace -$ cosmic-time
orthogonal foliations $ {\cal N}_3$ homothetic to $N_3$, the
metric on $M_4$ gets the well-known Friedmann-Robertson-Walker
(FRW) form
\begin{equation}
ds^2 \, = \, a^2 \, e^{2 \, f} \, dl_{N_3}^2 \, - \, (dt)^2 \,\, ,
\end{equation}
where $a$ is a scale parameter with dimension of length and the
modified metric function $f\, : \, {\rm{\bf R}} \rightarrow
{\rm{\bf R}}$ does actually express the primitive
\begin{equation}
f(t) \, = \, \int^t H(t') \, dt'
\end{equation}
of the celebrated Hubble function
\begin{equation}
H(t) \, \stackrel{\Delta}{=} \, \frac{e^{-f}}{a} \, \frac{d}{dt}
\left( a \, e^f \right) \, \equiv \frac{df}{dt}
\end{equation}

Consequently, with respect to (16) and (2), the dually related
pseudo-orthonormal bases $\lbrace {\mbox{\boldmath $\omega$}} ;
{\mbox{\boldmath $e$}} \rbrace$ are respectively given by the
concrete expressions
\begin{eqnarray}
& & \omega^1 = a e^f \cos(\theta) d \alpha \, , \, \omega^2 = a
e^f \sin(\theta) d \beta \, , \, \omega^3 = a e^f \, d \theta \, ,
\, \omega^4 = dt  \nonumber \\* & (a) & \omega^1 = a e^f d \alpha
\, , \, \omega^2 = a e^f d \beta \, , \, \omega^3 = a e^f d \theta
\, , \, \omega^4 = dt \nonumber \\* & & \omega^1 = a e^f
\cosh(\theta) d \alpha \, , \, \omega^2 = a e^f \sinh(\theta) d
\beta \, , \, \omega^3 = a e^f d \theta \, , \, \omega^4 = dt
\nonumber \\*
 & & e_1 =
\frac{e^{-f}}{a} {\rm sec} (\theta) \partial_{\alpha} \, , \, e_2 =
\frac{e^{-f}}{a} {\rm cosec} (\theta) \partial_{\beta } \, , \, e_3
= \frac{e^{-f}}{a} \partial_{\theta} \, , \, e_4 =
\partial_4
\nonumber \\*
& (b) & e_1 =
\frac{e^{-f}}{a}
\partial_{\alpha} \, , \, e_2 = \frac{e^{-f}}{a}
\partial_{\beta} \, , \, e_3 = \frac{e^{-f}}{a}
\partial_{\theta} \, , \, e_4 = \partial_t
\nonumber \\*
& & e_1 = \frac{e^{-f}}{a} {\rm sech} (\theta)
\partial_{\alpha} \, , \, e_2 = \frac{e^{-f}}{a} {\rm cosech}
(\theta) \partial_{\beta} \, , \, e_3 = \frac{e^{-f}}{a}
\partial_{\theta} \, , \, e_4 = \partial_t
\end{eqnarray}
which move the exterior-forms formalism, through the Cartan's
Equations
\begin{eqnarray}
&(a)& d \omega^a = \Gamma^{a}_{\; bc} \, \omega^b \wedge \omega^c
\, , \; {\rm without \; torsion} \, , \nonumber \\* &(b)& {\rm{\bf
R}}_{ab} = d \Gamma_{ab} \, + \, \Gamma_{ac} \wedge \Gamma^{c}_{\;
b} \, \, ,
\end{eqnarray}
where
\[
\Gamma_{ab} \, = \, \eta_{ad} \, \Gamma^{d}_{\; b} \, = \,
\Gamma_{abc} \, \omega^c \, \, ,
\]
all the way down to the essential components $R_{abcd}$ of the
curvature 2-forms,
\[
{\rm{\bf R}}_{ab} \, = \, \frac{1}{2} \,R_{abcd} \, \omega^c
\wedge \omega^d \, \, ,
\]
namely,
\begin{eqnarray}
&(a)& R_{1212} \, = \, R_{1313} \, = \, R_{2323} \, = \, \left(
f_{|4} \right)^2 \, + \, \frac{k}{a^2} e^{-2 \, f} \, , \,
\nonumber \\* &(b)& R_{1414} \, = \, R_{2424} \, = \, R_{3434} \,
= \, - \, \left[ f_{|44} \, + \, \left( f_{|4} \right)^2 \right]
\, .
\end{eqnarray}
Thus, the Ricci tensor gets no off-diagonal components and
therefore it reads
\begin{eqnarray}
&(a)& R_{\alpha \beta} \, = \, \left[ f_{|44} \, + \, 3 \left(
f_{|4} \right)^2 \, + \, \frac{2 \, k}{a^2} \, e^{-2 \, f} \right]
\, \delta_{\alpha \beta} \, \, , \nonumber \\* &(b)& R_{44} \, =
\, -3 \, \left[ f_{|44} \, + \, \left( f_{|4} \right)^2 \right] \,
\, ,
\end{eqnarray}
where $\alpha, \beta = \overline{1,3}$, leading to the scalar
curvature
\begin{equation}
R \, = \, 6 \left[ f_{|44} \,+ \, 2 \left( f_{|4} \right)^2 \, +
\, \frac{k}{a^2} \, e^{-2 \, f} \right]
\end{equation}
and altogether to the algebraically essential components of the
Einstein tensor
\begin{eqnarray}
&(a)& G_{\alpha \beta} \, = \, - \left[ 2 \, f_{|44} \, + \, 3
\left( f_{|4} \right)^2 \, + \, \frac{k}{a^2}\, e^{-2 \, f}
\right] \, \delta_{\alpha \beta} \nonumber \\* &(b)& G_{44} \, =
\, 3 \left[ \left( f_{|4} \right)^2 \, + \, \frac{k}{a^2} \, e^{-2
\, f} \right]
\end{eqnarray}

As it can be noticed, for both $k \, = \, \lbrace 0,1 \rbrace$ the
$G_{44}$-component does always stand non-negative and that is an
important restriction, through the Einstein equation $G_{44} =
\kappa_0 \, T_{44}$, on the type of matter-sources that can be fit
(in the sense of an exact solution) into such geometries:
excepting the Minkowskian ($k=0$)-vacuum case, all the other
matter-sources $-$ if combined $-$ should have a positive total
energy density, $w \, = \, T_{44} > 0$, i.e. they should behave on
the whole, for $k=0$ or $1$, as {\it conventional} matter,
fulfilling the Hawking's {\it weak} energy condition, $T_{ab} \,
u^a u^b \geq 0$ (for any non-spacelike 4-vector ${\mbox{\boldmath
$u$}}$). On the contrary, and completely nontrivial from a
geometrodynamical perspective, in the hyperbolic case, $k=-1$, the
sign of $G_{44}$ and (together with it) the one of the (resulting)
energy density gets undefined, unless $f_{|4}=0$ when the Einstein
equations demand a deeply exotic kind of matter, of state-equation
\[
P \, = \, - \, \frac{1}{3} \, w \, , \; {\rm where} \; w \, = \,
- \, \frac{3/ \kappa_{0}}{a^2} \, ,
\]
i.e.
\[
P \, = \, \frac{1}{3} \, |w| \, \, ,
\]
which can be comprehended as a sort of {\it ghost}-black-body
radiation. Anyway, in the general situation, of evolving
(time-orthogonal) $H^3-$foliations, the geometrodyamics gets much
more involved, since it can accommodate some reasonably mixed
matter-sources, made both of ^^ ^^ ordinary'' and ^^ ^^ exotic''
matter. Thus, letting apart for the moment the well-known
conventional sources, such as the thermalized electromagnetic (or
other massless-field) radiation and the baryonic dust, we deal
with the case where the $(k = -1)-$FRW geometry is driven by
the spontaneously inner parity breaking massive scalar field,
$\chi$, alone. As all of the essential Einstein tensor components
are on-diagonal and only time-dependent, the source field $\chi$
can only be {\it coherent} and therefore, its conservative
stress-energy-momentum tensor does also become diagonal,
exhibiting the components $-$ subsequently derived from (12) $-$
\begin{eqnarray}
&(a)& T_{\alpha \beta} \, = \, - \, \frac{1}{2} \left[ - \left(
\chi_{|4} \right)^2 \, + \, 2 \, V( \chi) \, - \, \frac{3
\mu^4}{\lambda} \right] \, \delta_{\alpha \beta} \nonumber \\*
&(b)& T_{44} \, = \, \frac{1}{2} \left[ \left( \chi_{|4} \right)^2
\, + \, 2 \, V( \chi) \, - \, \frac{3 \mu^4}{\lambda} \right]
\end{eqnarray}
Hence, the whole set of ^^ ^^ quartically generalized''
Einstein-Gordon equations
\[
G_{ab}[f] \, = \, \kappa_0 \, T_{ab}[ \chi]
\]
immediately goes down to the following functionally 2-dimensional
nonlinear  differential system
\begin{eqnarray}
&(a)& 2 \, f_{|44} \, + \, 3 \left( f_{|4} \right)^2 \, - \,
\frac{e^{-2 \, f}}{a^2} \, = \, \frac{\kappa_0}{2} \left[ - \,
\left( \chi_{|4} \right)^2 \, + \, 2 \, V( \chi) \, - \, \frac{3
\mu^4}{\lambda} \right] \, \, , \nonumber \\* &(b)& 3 \left[
\left( f_{|4} \right)^2 \, - \, \frac{e^{-2 \, f}}{a^2} \right] \,
= \, \frac{\kappa_0}{2} \left[ \left( \chi_{|4} \right)^2 \, + \, 2 \,
V( \chi) \, - \, \frac{3 \mu^4}{\lambda} \right] \, \, ,
\end{eqnarray}
where the semi-classical potential $V$ is given by (13). We have
not included in (26) the generalized Gordon equation (10),
explicitly worked out for the considered $(k = -1)-$FRW
spacetime  dynamically sustained by the coherent massive scalar
$\chi$, since all three of them, i.e. (26.a, b) and (10), taken
together, are not functionally independent because of the twice
contracted Second Bianchi Identity
 \begin{equation}
 0 \, \equiv \, \left( R^{ab} \, - \, \frac{1}{2} \, g^{ab} \,R
 \right)_{ ; \, b} \, = \, G^{ab}_{; \, b} \, = \,\kappa_0 \, T^{ab}_{;
\, b}
 \, \Rightarrow \, T^{ab}_{; \, b} \, = \,0 \, ,
\end{equation}
i.e. if the energy-momentum tensor ${\mbox{\boldmath $T$}}$ is
correctly derived from the field Lagrangian density ${\cal L}$
then its 4-divergenceless property (the one of being {\it
conservative}) does accurately account for the field dynamics
prescribed by the Euler-Lagrange equations
\[
\frac{\delta {\cal L}}{\delta \chi} \, = \,0 \, \, .
\]
To put it shortly, the source-field equation (10), particularized
to the form
\begin{equation}
\chi_{|44} \, + \, 3 \, f_{|4} \, \chi_{|4} \, + \, 2 \, \mu^2 \,
\chi \, = \,- \, 3 \mu \sqrt{\frac{\lambda}{6}}\, \chi^2 \,- \,
\frac{\lambda}{6} \, \chi^3 \; ,
\end{equation}
must spring out from (26) just by taking first-order derivatives
and subsequently doing algebraic manipulations. Indeed, taking
the time-derivative of(26.b) it yields
\[
2 \left[ f_{|44} \, + \, \frac{e^{-2 \,f}}{a^2} \right] \, = \,
\frac{\kappa_0}{3} \, \frac{ \chi_{|4}}{f_{|4}} \left[ \chi_{|44}
\, + \, \frac{dV}{d \chi} \right]
\]
and, by insertion in (26.a), written as
\[
 2 \left[ f_{|44} \, + \, \frac{e^{-2 \,f}}{a^2} \right] \, +
 \, 3 \left[ \left( f_{|4} \right)^2 \, - \, \frac{e^{-2
 \,f}}{a^2} \right] \, = \, \frac{\kappa_0}{2} \left( \chi_{|4}
 \right)^2 \, + \, \kappa_0 \, V( \chi) \, - \, \kappa_0 \frac{3
 \mu^4}{2 \lambda} \, \, ,
\]
i.e., using (26.b),
\begin{eqnarray}
2 \left[ f_{|44} \, + \, \frac{e^{-2 \,f}}{a^2} \right] & + &
\frac{\kappa_0}{2} \left( \chi_{|4} \right)^2 \, + \, \kappa_0 \,
V( \chi) \, - \, \kappa_0 \frac{3 \mu^4}{2 \lambda} \nonumber \\*
= &  - & \frac{\kappa_0}{2} \left( \chi_{|4} \right)^2 \, + \,
\kappa_0 \, V( \chi) \, - \, \kappa_0 \frac{3 \mu^4}{2 \lambda} \;
, \nonumber
\end{eqnarray}
it gives
\[
\frac{1}{3} \, \frac{\chi_{|4}}{f_{|4}} \left[ \chi_{|44} \, + \,
 \frac{dV}{d \chi} \right] \, + \, \left( \chi_{|4} \right)^2 \, =
 \, 0 \, \, ,
\]
i.e.
\[
\chi_{|44} \, + \, 3 f_{|4} \, \chi_{|4} \, = \, - \,
\frac{dV}{d \chi} \, , \; \; {\rm with} \; \; \chi_{|4} \neq 0 \neq
f_{|4} \, \, ,
\]
where the last equation does exactly come to the Gordon one (28)
just by plugging in the $\chi$-derivative of the fourth-degree
polynomial potential (13).

Nevertheless, is good to know that we can play all three
equations, (26.a, b) and (28), since, in some concrete
calculations, the result might be got easier in some particular
combination of them, instead of working only with (26) as they
stand.

In the above given proof of the compatibility of Euler-Lagrange
equation (28) with the Einstein's ones (26), we have asked for
$f_{|4} \neq 0$ and $\chi \neq 0$. If $f_{|4} =  0$ then $f$
can be scaled to {\it zero} and the system (26) becomes
\begin{eqnarray}
\frac{1}{a^2} & = & \frac{\kappa_0}{2} \left( \chi_{|4} \right)^2
\, - \, \kappa_0 \,  V(\chi ) \, + \, \frac{3 \kappa_0 \mu^4}{2
\lambda} \nonumber
\\*
\frac{3}{a^2} & = & - \, \frac{\kappa_0}{2} \left( \chi_{|4}
\right)^2 \, - \, \kappa_0 \, V(\chi ) \, + \, \frac{3 \kappa_0
\mu^4}{2 \lambda} \nonumber
\end{eqnarray}
Subtracting the first equation from the second one, it yields
\[
\left( \chi_{|4} \right)^2 = \, - \, \frac{2/\kappa_0}{a^2} \; \;
\Rightarrow \; \; \chi = \chi_0 \pm i \, \sqrt{\frac{2}{\kappa_0}}
\, \frac{t}{a} \, ,
\]
so that, the massive scalar $\chi$ would be actually a {\it
ghost}; in addition, since $\chi_{|44} \equiv 0$ (in this case),
the nonlinear Gordon equation (28) just turns into the algebraic
equation
\[
\chi^2 \, + \, 3 \mu \sqrt{\frac{\lambda}{6}} \, \chi \, + \,
\frac{12 \mu^2}{\lambda} \, = \, 0 \, , \; {\rm with} \; \; \chi
\sim \frac{t}{a} \, , \; (\forall ) \, t \in {\rm{\bf R}} \, ,
\]
which obviously cannot be satisfied as $\chi$ is clearly
time-dependent. Hence, the $(f = const)-$particular case is
definitely forbidden for the considered matter-source.

\section{Maximally Symmetric Fixed Points}

The other particular case, $\chi_{|4} =0$, is by far of much
interest for it reveals the simplest $(k=-1)-$FRW spacetime
dynamics in the fixed points of the nonlinear Gordon equation for
the physical field $\chi$ left-over by the spontaneous breaking of
the discrete inner-symmetry $\phi \rightarrow - \phi$. The
shortest path to the solution(s) is paved by the observation that
for $\chi_{|4} =0$ the right-hand-side of the two Einstein
equations (26) gets the same and therefore, subtracting them, it
yields the modified metric function essential equation
\begin{equation}
f_{|44} \, + \, \frac{e^{-2f}}{a^2} \, = \, 0
\end{equation}
which can be inserted back into (26.a), with $\chi_{|4} =0$,
getting at once the same equation (26.b) (with $\chi_{|4} =0$).
Hence, among the three Einstein-(generalized) Gordon equations we
have just to solve, in this particular case, the very simple
system
\begin{eqnarray}
& (a) & \left( f_{|4} \right)^2 - \, \frac{e^{-2f}}{a^2} \, = \,
\frac{\kappa_0 }{3} \, V(\chi ) \, - \, \frac{\kappa_0 \mu^4}{2
\lambda} \nonumber \\* & (b) & \chi \left[ 2 \mu^2 \, + \, 3 \mu
\, \sqrt{\frac{\lambda}{6}} \, \chi \, + \, \frac{\lambda}{6} \,
\chi^2 \right] = \, 0
\end{eqnarray}
where $V(\chi )$ is given by (13), i.e.
\begin{equation}
V(\chi ) \, = \, \frac{\lambda}{24} \, \chi^2 \left[ \chi^2 \, +
\, 4 \mu \, \sqrt{\frac{6}{\lambda}} \, \chi \, + \, \frac{24
\mu^2}{\lambda} \right]
\end{equation}
Ordered by their magnitudes, the roots of (30.b) $-$ meaning the
matter-field fixed-point values $-$ do respectively read
\begin{equation}
\chi_L \, = \, - \, 2 \mu \, \sqrt{\frac{6}{\lambda}} \, , \;
\chi_M \, = \, - \, \mu \, \sqrt{\frac{6}{\lambda}} \, , \; \chi_R
\, = \, 0 \, ,
\end{equation}
where the indices $L, \, M, \, R$ come from ^^ ^^ left, Milne,
right'', respectively. By their ^^ ^^ turn by turn'' insertion in
(31), one immediately gets the corresponding values of the
semi-classical (4-nominal self-interaction) potential, namely
\begin{equation}
V_L \, = \, 0 \, , \; V_M \, = \, \frac{3 \mu^4}{2 \lambda} \, ,
\; V_R \, = \, 0 \, ,
\end{equation}
so that, the nonlinear first-order differential equation (30.a) of
the metric function $f$ does only take the following two
particular forms
\begin{eqnarray}
& (a) & \left( f_{|4} \right)^2 - \, \frac{e^{-2f}}{a^2} \, = \, -
\, \frac{\kappa_0 \mu^4 }{2 \lambda} \, , \; {\rm for } \; \; V_L
= V_R =0  \, , \nonumber \\* & (b) & \left( f_{|4} \right)^2 - \,
\frac{e^{-2f}}{a^2} \, = \, 0 \, , \; {\rm for } \; \; V_M = \,
\frac{3 \mu^4}{2 \lambda} \, ,
\end{eqnarray}
which will be correspondingly leading to the only two types of
$k=-1$ homogeneous and isotropic universes that can accommodate
and are also physically supported by the massive, $m_{\chi} =
\sqrt{2} \, \mu$, real scalar field $\chi$ in any of its three
4-dimensionally constant main states. It is worth noticing that
without integrating the equation (34) we can already say, by the
help of equation (29), what the two generated spacetimes are.
Indeed, in the simpler case (34.b), because of (29), it also
results that
\begin{equation}
f_{|44} \, + \, \left( f_{|4} \right)^2 \, = \, 0
\end{equation}
and thus, having a look at the components (21) of the curvature
tensor, we instantly realize that all of them vanish. Hence, the
spacetime corresponding to (34.b), supported by the
static physical field $\chi_M = - \, \mu \,
\sqrt{6/ \lambda}$, is {\it flat}, being basically (geometrically) the
Minkowski spacetime. However, especially from a cosmological
perspective, the difference is that this spacetime is patched in a
different coordinate-system, namely the Milne's one, which sharply
presents the evolution of the $H^3-$spacelike-foliation, instead
of the static picture of the Cartesian ${\rm{\bf R}}^3$-foliations
(of constant Minkowskian time, $x^4 =t$). Concerning the other
$(k=-1)-$FRW model, the one related to the equation (34.a), we
get, using again the essential equation (29), that
\begin{equation}
f_{|44} \, + \, \left( f_{|4} \right)^2 \, = \, - \,
\frac{\kappa_0 \mu^4}{2 \lambda}
\end{equation}
and so, inserting it, together with (34.a), in the expressions
(21) of the Riemann tensor components and in the expression (23),
with $k=-1$, of the scalar curvature, it yields
\begin{eqnarray}
& (a) & R_{\alpha \beta \alpha \beta} \, = \, - \, R_{\alpha 4
\alpha 4 } \, = - \, \frac{\kappa_0 \mu^4}{2 \lambda} \, , \;
\alpha = \overline{1,3} \, , \; \beta \neq \alpha \, , \; \beta =
\overline{1,3} \, , \nonumber \\* & (b) & R \, = \, - \, 12 \left(
\frac{\kappa_0 \mu^4}{2 \lambda} \right) ,
\end{eqnarray}
which clearly points out that the solution to the ^^ ^^ basic''
equation (34.a) does surely sustain a $(k=-1)-$FRW Universe of
constant negative (4D) curvature and that can only be the anti-de
Sitter spacetime. Although investigating it, and some of its
possibly observable cosmological consequences, mostly related to
the spontaneous breaking of the field-reflection inner symmetry,
is the main goal of the present paper, we would like (first) to
make some remarks on the model described by the solution of
(34.b), better known as the Milne Universe, and to comment a bit
its linear stability within the context of the coherent linear
perturbations of the massive source-field $\chi$ around its
fixed-point value $\chi_M$.

\section{Milne Spacetime Coherent Perturbations}

As it is almost obvious, because of its very simple form, the
equation (34.b) can be immediately written as
\begin{equation}
\left( \frac{dS}{dt} \right)^2 \, = \, 1 \, ,
\end{equation}
where
\begin{equation}
S \, = \, a \, e^f
\end{equation}
is the ^^ ^^ standard'' cosmological scale function $-$ casting
the FRW-metric into the ^^ ^^ history making'' form
\begin{equation}
ds^2 \, = \, S^2(t) \, dl_{N_3}^2 \, - \, (dt)^2 \, ,
\end{equation}
(where $N_3$ is one of the $S^3$, ${\rm{\bf R}}^3$, $H^3$
manifolds) $-$ and so, independently of both the sign we choose
for the solution and the corresponding integration constant, the
scale function is basically reading
\begin{equation}
S(t) \, = \, |t| \, ,
\end{equation}
so that, in $(k=-1)-$spherical (physically dimensionless)
coordinates $\lbrace r, \theta , \varphi \rbrace$ on the upper
wing (let's say) of $H^3$, the {\it generic} metric (40) does
explicitly turn into the one of the Milne Universe
\begin{equation}
ds_{Mln}^2 \, = \, t^2 \left[ \frac{(dr)^2}{1+r^2} \, + \, r^2 \,
d \Omega^2 \right] \, - \, (dt)^2 \, ,
\end{equation}
where
\begin{equation}
d \Omega^2 \, = \, (d \theta)^2 \, + \, \sin^2 \theta \, (d \varphi)^2
\end{equation}
is the well-known metric on the unit sphere $S^2$. Since, as we have
already
shown, this spacetime is properly {\it flat}, there must be a globally
defined local coordinates transformation that takes the Minkowskian
metric,
\begin{equation}
d s^2_{Mnk} \, = \, \left( d R \right)^2 \, + \, R^2\, d \Omega^2 \,
- \, \left( d T \right)^2 \, \, ,
\end{equation}
into the Milne's one, (42), and vice versa. In this respect, as
the angular $( \theta, \varphi)-$coordinates are the same (in the two
metrics and on the two manifolds), it is quite obvious that the
usual radial coordinate $R$ is given in terms of the Milne
coordinates ^^ ^^ $r,t$'' by the simple relation
\begin{equation}
R(r,t) \, = \, r \, |t| \, , \; {\rm with} \; r \geq \, 0 \, .
\end{equation}
Then, the Minkowskian time $T$ must also be a function of the two
coordinates
$(r,t)$, i.e.
\begin{equation}
T \, = \, T(r , t ) \, ,
\end{equation}
such that, plugging in (44) the square of its differential and the one
of (45),
it should equate the two (1+1)-metrics, i.e.
\begin{equation}
\left( d R \right)^2 \, - \, \left( d T \right)^2 \, = \,
\frac{t^2}{1 \, + \, r^2} \, (d r)^2 \, - \, (d t)^2 \, \, ,
\end{equation}
which concretely comes to the self-embedding condition
\begin{equation}
\left( |t| \, dr \, + \, r \, \frac{d |t|}{dt} \, dt \right)^2
- \left( \frac{\partial T}{\partial r} \, dr \, + \,
\frac{\partial T}{\partial t} \, dt \right)^2 = \,
\frac{t^2}{1 + r^2} \, (dr)^2 \, - \, (dt)^2 \, ,
\end{equation}
leading therefore to the following set of equations that (46) must
satisfy:
\begin{eqnarray}
&(a)& \left( \frac{\partial T}{\partial r} \right)^2 \, = \,
\frac{t^2 \, r^2}{1 + r^2} \nonumber \\* &(b)& \frac{\partial
T}{\partial r} \, \frac{\partial T}{\partial t} = \, r \, t
\nonumber \\* &(c)& \left( \frac{\partial T}{\partial t} \right)^2
\, = \, 1 + r^2
\end{eqnarray}
Hence, setting to zero the integration constants, the two branches
that spring out from the equations (49.a, c), compatible with (49.b),
are respectively defined by
\begin{eqnarray}
&(a)& \frac{\partial T}{\partial t} = - \, \sqrt{1 + r^2} \,
\Rightarrow \, \frac{\partial T}{\partial r} = - \frac{r
t}{\sqrt{1 + r^2}} \, \Rightarrow \, T = \, - \, t \, \sqrt{1 +
r^2} \, \, , \nonumber \\* &(b)& \frac{\partial T}{\partial t} =
\, \sqrt{1 + r^2} \, \Rightarrow \, \frac{\partial T}{\partial r}
= \, \frac{r t}{\sqrt{1 + r^2}} \, \Rightarrow \, T = \, t \,
\sqrt{1 + r^2} \, \, ,
\end{eqnarray}
where only (50.b) preserves the same orientation on the considered
manifold,
i.e.
\begin{equation}
\left|
\begin{array}{l}
\frac{\partial R}{\partial r} \; \; \; \; \frac{\partial R}{\partial t}
\\
\frac{\partial T}{\partial r} \; \; \; \; \frac{\partial T}{\partial t}
\end{array}
\right| \, = \, \frac{|t|}{\sqrt{1 + r^2}} > 0 \, , \; \forall \;
t \in {\rm{\bf R}} - \lbrace 0 \rbrace \, .
\end{equation}
Thus, collecting the two results (45) and (50.b), we have actually
got the celebrated {\it orientation-preserving} Milne
transformation,
\begin{eqnarray}
&(a)& R \, = \, r \, |t| \, \, , \nonumber \\* &(b)& T \, = \, t
\, \sqrt{1 + r^2} \, ,
\end{eqnarray}
where $r \geq 0$ and $t \in {\rm{\bf R}}$, which
casts (44) into the form (42) and, with a bit of care (again,
about orientation), we can immediately write down, from (52), its
(proper) {\it inverse}
\begin{eqnarray}
&(a)& r \, = \, \frac{r}{\sqrt{T^2 - R^2}} \nonumber \\* &(b)& t
\, = \, sgn(T) \, \sqrt{T^2 - R^2} \, \, ,
\end{eqnarray}
with $T^2 - R^2 \geq \, 0 \, , \, \,  R \geq \, 0 \, , \, \, T \in
\textbf{R}$, which actually defines the Milne coordinates $(r,t)$
in terms of the Minkowskian ones $(R,T)$ and accordingly turns
(42) into (44). Although the bulk of the Milne's Universe
properties and structure is very well known since quite long ago,
its connection to the spacetime geometry erected by the
^^ ^^ remnant'' massive scalar field $\chi$, ^^ ^^ after'' the
$\textbf{Z}_{2}-$invariance got spontaneously broken, has not been
investigated to a large extent in the literature. It might be
shedding a brand new light on the real (physical) significance of
the Milne Universe $-$ previously considered just as a ^^ ^^ toy
model'' $-$ and on a seemingly not yet extensively explored bunch
of cosmological implications of the decaying remnant-field blown
up Milne bubbles in the observable Universe. That is why is worth
studying their behaviour at least with respect to the linear
perturbations of the source-field $\chi$ around its static value
$\chi_M$.

Hence, let us consider the coherently evolving small field-perturbation
$\psi$, such that
\begin{equation}
\chi = \chi_M \, + \, \psi \, , \; {\rm where} \; \; | \psi | << |
\chi_M | = \sqrt{\frac{6}{\lambda}} \, \, ,
\end{equation}
whose dynamics is effectively controlled by (28), which concretely
becomes
\begin{equation}
\frac{d^2 \, \psi}{d t^2} \, + \, \frac{3}{t} \, \frac{d \psi}{d t} \,
- \, \mu^2 \, \psi \, = \, 0 \, \, ,
\end{equation}
once the expression
\begin{equation}
f \, = \, \ln \left( \frac{|t|}{a} \right) \, \, ,
\end{equation}
derived from (39) and(41), and the identity
\begin{equation}
\frac{\lambda}{6} \, \chi_M^2 \, + \, 3 \mu \sqrt{\frac{\lambda}{6}} \,
\chi_M
\, + \, 2 \mu^2 \, \equiv \, 0 \, \, ,
\end{equation}
stated by (30.b), are taken into account. Because of the
matter-field fluctuations encoded in (55), the initial Milne
background starts exhibiting metric perturbations which no longer
fulfill the unperturbed equations (34.b) and (35); instead, their
evolution is controlled by the two Einstein equations (26.a, b).
Fortunately, since the last two terms on the right hand side of
(26) are exactly the same, one gets a single essential equation
for the perturbed modified metric function, $f$, namely
\begin{equation}
f_{|44} \, + \, \frac{e^{-2 \, f}}{a^2} \, = \, - \,
\frac{\kappa_0}{2}\, \left( \chi_{|4} \right)^2 \, ,
\end{equation}
just by subtracting (26.b) from (26.a). As it can be noticed,
although one can reduce the order of differentiability in (58), by
the well-known substitutions
\begin{equation}
p = \frac{df}{dt} \, , \, dt = \frac{df}{p} \; \Rightarrow \;
\frac{d}{dt} = p \, \frac{d}{df} \, , \; q = p^2 \, \, ,
\end{equation}
casting it into the form of the linear, non homogeneous, first-order
differential equation
\begin{equation}
\frac{d q}{d f} \, + \, \kappa_0 \, \left( \frac{d \psi}{d f} \right)^2
q \, = \, - \, \frac{2}{a^2}\, e^{-2 \, f}
\end{equation}
$- \; \; \kappa_0 \left( \chi_{|4} \right)^2 = \kappa_0 \, p^2
\left( \frac{d \chi}{d f} \right)^2 = \kappa_0 \left( \frac{d
\psi}{d f} \right)^2 q$ because of (54) and (59) $-$ whose
solution reads
\[
q = e^{- \, \kappa_0 \, \int^f \left( \frac{d \psi}{d \xi}
\right)^2 \, d
\xi} \,  \left[ {\cal C} \, - \, \frac{2}{a^2} \int^f \, e^{-2 \left[
\xi - \frac{\kappa_0}{2} \int^{\xi} \left( \frac{d \psi}{d \eta}
\right)^2 \, d
\eta \right]} \, d \xi \right] \, ,
\]
where ${\cal C} = const. \, \geq \, 0 \,$, so that
\[
\frac{d f}{d t} \, = \pm \, e^{- \frac{\kappa_0}{2} \int^f \left(
\frac{d \psi}{d \xi} \right)^2 \, d \xi} \, \left[{\cal C} \, - \,
\frac{2}{a^2} \int^f \, e^{-2 \left[ \xi - \frac{\kappa_0}{2}
\int^{\xi} \left( \frac{d \psi}{d \eta} \right)^2 \, d \eta
\right]} \, d \xi \right]^{1/2}
\]
and therefore
\[
t \, = \, t_0 \, \pm \, \int \frac{e^{\frac{\kappa_0}{2} \int^f
\left( \frac{d \psi}{d \xi} \right)^2 \, d \xi}\, \, df} {\left[
{\cal C} \, - \, \frac{2}{a^2} \int^f \, e^{-2 \left[ \xi -
\frac{\kappa_0}{2} \int^{\xi} \left( \frac{d \psi}{d \eta}
\right)^2 \, d \eta \right]} \, d \xi \right]^{1/2}} \, , \, t_0
\in {\rm{\bf R}} \, ,
\]
one cannot actually get an explicit closed-form solution quite
because of the matter-source contribution $\frac{\kappa_0}{2}
\left( \frac{d \psi}{d f} \right)^2$ which, in principle, should
be worked out from the awkward (although linear) second-order
differential equation of the linear field-fluctuation around
$\chi_M$
\begin{equation}
\frac{d^2 \, \psi}{d f^2} \, + \, \left[ 3 \, + \, \frac{d}{df}
\left( \ln \sqrt{q}  \, \right) \right] \frac{d \psi}{d f} \, - \,
\frac{\mu^2}{q} \, \psi \, = \, 0 \, \, ,
\end{equation}
representing the image of (55), softly generalized as
\[
\frac{d^2 \, \psi}{d t^2} \, + \, 3 \, \frac{d f}{d t} \, \frac{d
\psi}{d t} \, - \, \mu^2 \, \psi \, = \, 0 \, \, ,
\]
under the whole set of substitutions (59). Hence, even for small
perturbations around the Milne value of the physical field $\chi$,
it is clearly unlikely to get an exact solution $\lbrace \psi\, ,
\, q \rbrace (f) \, : \textbf{R} \rightarrow \lbrace \textbf{R} \,
, \, \textbf{R}_{+} \rbrace$, i.e., by (59), $\lbrace \psi \, , \,
f \rbrace(t) \, : \textbf{R} \rightarrow \textbf{R}$, to the
highly nonlinearly coupled differential equations (60) and (61)
and therefore, analytically, the best one can do, at least for
tackling the subject, is to consider the linear fluctuation $h$ in
the modified metric function $f$, around its Milne form (56), i.e.
\begin{equation}
f \, = \, \ln \left| \frac{t}{a} \right| \, + \, h \, \, ,
{\rm with} \; e^h \, \cong \, 1 + h
\end{equation}
and to subsequently linearize the essential (inhomogeneous)
Einstein equation (58), getting the far much simpler form
\begin{equation}
\frac{d^2 h}{d t^2} \, - \, \frac{2}{t^2} \, h \, = \, - \,
\frac{\kappa_0}{2}\, \left( \frac{d \psi}{d t} \right)^2 \, \, ,
\end{equation}
where the source-term on the right hand side is going to be
evaluated, in the first-order approximation, by integrating the
fluctuating remnant field equation (55). In this respect, setting
\begin{equation}
\tau = \mu \, t \; \; {\rm and} \; \; \psi \, = \, \tau^{\nu} \,
U(\tau) \, , \, \, \nu \in {\rm{\bf R}} \, \, ,
\end{equation}
it yields the differential equation
\begin{equation}
\frac{d^2 U}{d \tau^2} \, + \, \frac{2 \nu + 3}{\tau} \,
\frac{d U} {d \tau} \, - \, \left(1 \, - \,
\frac{\nu (\nu + 2)}{\tau^2} \right) \, U \, = \, 0 \, \, ,
\end{equation}
for the new function $U$, so that, for
\begin{equation}
\nu \, = \, -1 \, \, ,
\end{equation}
it concretely gets the form of the ^^ ^^ {\it modified} Bessel
functions'' equation,
\begin{equation}
\frac{d^2 U}{d \tau^2} \, + \, \frac{1}{\tau} \,
\frac{d U} {d \tau} \, - \left(1 \, + \,
\frac{1}{\tau^2} \right) U \, = \, 0 \, \, ,
\end{equation}
and therefore, referring to the perturbation field $\psi$ in (64)
with (66), one instantly gets the two (functionally) linearly
independent modes
\begin{equation}
\psi_{+}(t) \, = \, \frac{{\cal N}_{+}}{\mu t} \, I_{1}(\mu \, t)
\, \, , \, \, \psi_{-}(t) \, = \, \frac{{\cal N}_{-}}{\mu t} \,
K_{1}(\mu \, t) \, \, ,
\end{equation}
where the real constants ${\cal N}_{\pm} \, \, -$ of
renormalization dimension $D = 1 \, \, -$ set the amplitude of the
physical field fluctuation at some reference-moment ^^ ^^ $t_0$''.
In spite of the way it heads into the future (timelike) infinity,
decaying extremely fast, the mode $\psi_{-}$ is strongly singular
at the Milne-time origin, $t=0$, and takes very large values
around it, so that, at such early epochs, it hardly can be
considered as a {\it small} perturbation that should fulfill the
requirement $ | \psi_{-}(0 < \mu t << 1)| << \mu \sqrt{6/ \lambda}
= |\chi_{M}|$. Therefore, only the $(t=0)-$nonsingular growing
mode $\psi_{+}$ contributes to the source term in (63) and drives
the evolution of the modified metric function perturbation, $h$.
With respect to the dimensionless time-like parameter $\tau$
(defined in (64)) the equation (63) does simply read
\begin{equation}
\frac{d^2 h}{d \tau^2} \, - \, \frac{2}{\tau^2} \, h \, = \,
- \, \frac{\kappa_0}{2} \left( \frac{d \psi_{+}}{d \tau} \right)^2
\end{equation}
and we work out its general solution(s) by starting with the function
substitution
\begin{equation}
h \, = \, \tau^{\gamma} \, F(\tau) \, \, ,
\end{equation}
which turns (69) into the inhomogeneous Euler equation,
for the function $F$,
\begin{equation}
\frac{d^2 F}{d \tau^2} \, + \, \frac{2 \gamma}{\tau} \, \frac{d
F}{d \tau} \, + \, \frac{\gamma (\gamma - 1) - 2}{\tau^2} \, F \,
= \, - \, \frac{\kappa_0}{2} \, \left( \frac{d \psi_{+}}{d \tau}
\right)^2 \tau^{- \gamma} \, .
\end{equation}
As it can be noticed regarding the term in F, this can be
nontrivially vanished by taking $\gamma$ as each of the two roots
of the very simple 2-nd degree equation
\begin{equation}
\gamma^2 \, - \, \gamma \, - \, 2 \, = \, 0 \, \Rightarrow \,
\gamma = \lbrace - \, 1 \, , \, 2 \rbrace
\end{equation}
and therefore we get $-$ in principle $-$ two branches of solutions,
respectively stated by the mathematical relations
\begin{eqnarray}
&(a)& h_{-} \, = \, \frac{1}{\tau} \, F_{-}( \tau) \, \, , \, \,
\frac{d F_{-}}{d \tau} \, = \, G_{-}( \tau) \, \, , \nonumber
\\* &(b)& \frac{d G_{-}}{d \tau} \, - \, \frac{2}{\tau} \,
G_{-} \, = \, - \, \frac{\kappa_0}{2} \, \left( \frac{d
\psi_{+}}{d \tau} \right)^2 \, \tau \, , \, \, {\rm for} \, \gamma
\, = \, - 1 \, \, ; \nonumber \\* &(c)& h_{+} \, = \, \tau^2 \,
F_{+}( \tau) \, \, , \, \, \frac{d F_{+}}{d \tau} \, = \, G_{+}(
\tau) \, \, , \nonumber \\* &(d)& \frac{d G_{+}}{d \tau} \, + \,
\frac{4}{\tau} \, G_{+} \, = \, - \, \frac{\kappa_0}{2} \,
\frac{1}{\tau^2} \left( \frac{d \psi_{+}}{d \tau} \right)^2\, , \,
\, {\rm for} \, \gamma \, = \, 2 \, \, .
\end{eqnarray}
Thus, by (73.a, c), we have reduced the second-order differential
equation(69)
to the inhomogeneous first-order differential equations (73.b, d) whose
solutions can be easily derived as
\begin{eqnarray}
&(a)& G_{-}( \tau) \, = \, G_{-}^{0} \tau^2 \, - \,
\frac{\kappa_0}{2} \, \tau^2 \, \int^{\tau} \frac{d s}{s} \left(
\frac{d \psi_{+}}{d s} \right)^2 \, , \, G_{-}^{0} = cst \in
{\rm{\bf R}} \, , \nonumber \\* &(b)& G_{+}( \tau) \, = \,
\frac{G_{+}^{0}}{\tau^4} \, - \, \frac{\kappa_0}{2} \,
\frac{1}{\tau^4} \, \int^{\tau} s^2 \left( \frac{d \psi_{+}}{d s}
\right)^2 d s \, , \, G_{+}^{0} = cst \in {\rm{\bf R}}\, ,
\end{eqnarray}
where $\psi_{+}$ is given by (68), with $ \mu t$ replaced by $s$, i.e.
\begin{equation}
\psi_{+}(s) = \frac{{\cal N}_{+}}{s} \, I_{1}(s) \, \Rightarrow \,
\left \lbrace
\begin{array}{l}
\frac{d \psi_{+}}{d s} = \frac{{\cal N}_{+}}{s} \,
\left[ \frac{d I_{1}}{d s} \, - \, \frac{I_{1}}{s} \right] \, \, , \\
\\ \frac{d I_{1}}{d s} = \frac{1}{2}
\left[ I_{0}(s) \, + \, I_{2}(s) \right] \, \, . \\
\end{array} \right.
\end{equation}

Making (a respective) use of (73.a, c), once (74.a, b) have been
gotten, we come to the concrete expression(s)
of the $-$ ^^ ^^ formally'' {\it two} $-$ (most) general solutions,
\begin{eqnarray}
&(a)& h_{-}( \tau) = F_{-}^{0} \tau^{-1} + \frac{1}{3} G_{-}^{0}
\tau^2 - \frac{\kappa_0}{2} \, \tau^{-1} \int \tau^2 \left(
\int^{\tau} \frac{d s}{s} \left( \frac{d \psi_{+}}{d s} \right)^2
\right) d \tau \, , \nonumber \\* &(b)& h_{+}( \tau) = -
\frac{G_{+}^{0}}{3} \tau^{-1} + F_{+}^{0} \tau^2 -
\frac{\kappa_0}{2} \, \tau^2 \, \int \frac{d \tau}{\tau^4} \left(
\int^{\tau} s^2 \left( \frac{d \psi_{+}}{d s} \right)^2 \, d s
\right) ,
\end{eqnarray}
with $F_{\pm}^{0} = cst \in {\rm{\bf R}}$,
of the linearized Einstein-Gordon equation (69) $\Leftrightarrow$
(63), which describes, envisaging(62), the coherent Milne's
Universe metric fluctuations induced by an initially small
perturbation $-$
\[
| \psi_{+}(0) | \, = \, \frac{1}{2} | {\cal N}_{+} | \, << \, \mu
\sqrt{\frac{6}{\lambda}} \, = \, | \chi_{M} | \, \, ,
\]
right on the singular, but {\it free} of ^^ ^^ real'' geometrical
singularities, $\lbrace t = 0 \rbrace-$foliation $-$ in the
physical field $\chi$, leftover around $\chi_M$ by the spontaneous
breaking of the inner parity invariance. Nevertheless, in a
straight computational manner, it is not quite trivial to realize
that the {\it two} branches {\it do} actually {\it coincide}.

The difference in the way they look is only apparent and comes
from an additional part-by-part integral that has been subtly
performed $-$ ^^ ^^ by itself'', actually $-$ in the switch from
(73.a, b) to (73.c,d). Indeed, one can notice first that
\begin{equation}
\int^{\tau} \frac{d s}{s} \left( \frac{d \psi_{+}}{d s} \right)^2
\, = \, \int^{\tau} \frac{d s}{s^3} \, s^2 \left( \frac{d
\psi_{+}}{d s} \right)^2 = \int^{\tau} \frac{1}{s^3} \left[ s^2
\left( \frac{d \psi_{+}}{d s} \right)^2 \right] d s
\end{equation}
and setting
\begin{eqnarray}
& & u \, = \, \frac{1}{s^3} \; \; , \; \; d v \, = \, s^2 \left(
\frac{d \psi_{+}}{d s} \right)^2 d s \, \, , \nonumber \\* & & d u
\, = \, - \, \frac{3 d s}{s^4} \; \; , \; \; v \, = \, \int^{s}
\xi^2 \left( \frac{d \psi_{+}}{d \xi} \right)^2 d \xi \, \, ,
\end{eqnarray}
it yields (integrating ^^ ^^ by parts'')
\begin{equation}
\int \frac{d \tau}{\tau} \left( \frac{d \psi_{+}}{d \tau}
\right)^2 \, = \, \frac{1}{\tau^3} \, \int s^2 \left( \frac{d
\psi_{+}}{d s} \right)^2 d s \, + \, 3 \, \int \frac{d
\tau}{\tau^4} \left( \int^{\tau} s^2 \left( \frac{d \psi_{+}}{d s}
\right)^2 d s \right) \, \, .
\end{equation}
That is to be used in the part-by-part integral
\[
\int \tau^2 \left( \int^{\tau} \frac{d s}{s}
\left( \frac{d \psi_{+}}{d s} \right)^2 \right) d \tau =
\frac{\tau^3}{3} \int^{\tau} \frac{d s}{s} \left(
\frac{d \psi_{+}}{d s} \right)^2 \, - \,
\frac{1}{3} \int \tau^2
\left( \frac{\psi_{+}}{d \tau} \right)^2 d \tau \, \, ,
\]
i.e.
\begin{eqnarray}
\int \tau^2 \left( \int^{\tau} \frac{d s}{s} \left( \frac{d
\psi_{+}}{d s} \right)^2 \right) d \tau \, &=& \frac{1}{3}
\int^{\tau} s^2 \left( \frac{d \psi_{+}}{d s} \right)^2 d s \, +
\nonumber \\* & & + \, \tau^3 \, \int \frac{d \tau}{\tau^4} \left(
\int^{\tau} s^2 \left( \frac{d \psi_{+}}{d s} \right)^2 d s
\right) \, - \nonumber
\\* & & - \, \frac{1}{3} \int^{\tau} s^2 \left( \frac{d
\psi_{+}}{d s} \right)^2 d s \, = \nonumber \\* &=& \tau^3 \int
\frac{d \tau}{\tau^4} \left( \int^{\tau} s^2 \left( \frac{d
\psi_{+}}{d s} \right)^2 d s \right) ,
\end{eqnarray}
explicitly stating that
\begin{equation}
\frac{1}{\tau}\, \int \tau^2
\left( \int^{\tau} \frac{d s}{s}
\left( \frac{d \psi{+}}{d s} \right)^2 \right) d \tau \, = \,
\tau^2 \, \int \frac{d \tau}{\tau^4}
\left( \int^{\tau} s^2
\left( \frac{d \psi_{+}}{d s} \right)^2 d s \right)
\end{equation}
and therefore, with
\begin{eqnarray}
& & - \, \frac{G_{+}^0}{3} \, = \, F_{-}^0 \, = \, F_0 \, \, ,
\nonumber \\* & & F_{+}^0 \, = \, \frac{G_{-}^0}{3} \, = \, - \,
\frac{1}{6} \, G_0 \, \, , \nonumber
\end{eqnarray}
it has been entirely proven that the two seemingly different
branches (76) are actually the same, being subsequently described
by the modified metric function perturbation
\begin{equation}
h \, = \, \frac{F_0}{\tau} \, - \, \frac{G_0}{6} \, \tau^2 \, - \,
\frac{\kappa_0}{2} \, \tau^{-1} \int \tau^2 \left( \int^{\tau}
\frac{d s}{s} \left( \frac{d \psi_{+}}{d s} \right)^2 \right) d
\tau \, \, .
\end{equation}
With that and (62), the {\it proper} scale function (39) becomes
\[
 S( \tau) \, = \, \frac{F_0}{\mu} \, + \, \frac{\tau}{\mu}
 \left[ 1 \, - \, \frac{G_0}{6} \, \tau^2 \, - \,
\frac{\kappa_0}{2} \, \tau^{-1} \, \int \tau^2 \left( \int^{\tau}
\frac{d s}{s} \left( \frac{d \psi_{+}}{d s} \right)^2 \right) d
\tau \right] ,
\]
so that, the seemingly divergent term $F_0 / \tau$, in (82),
brings nothing more than a {\it constant} (universal) shift in the
Milne's cosmic-time, making no contribution at all to the
curvature disturbances produced by the source-field (linear)
fluctuations $\psi_{+}$. Hence, it simply can be dropped away just
by setting $F_0 = 0$.

Quite on the contrary, the {\it seemingly arbitrary} constant
$G_0$ makes an important contribution to the curvature of the
coherently $\psi_{+}-$ perturbed Milne Universe, mostly with
respect to its {\it stability}.

To give the details of this matter, let us first work out the
curvature perturbations straight from the relations (82), (62) and
(21, with $k = - \, 1$). It primarily results
\begin{eqnarray}
R_{\alpha \beta \alpha \beta } & = & \frac{2 \mu^2}{\tau^2} \left[
\tau \, \frac{dh}{d \tau} + h \right] = \mu^2 \, \frac{2}{\tau^2}
\, \frac{d \;}{d \tau} ( \tau h) \, , \; \; \alpha , \beta =
\overline{1,3} \, , \nonumber \\* R_{\alpha 4 \alpha 4} & = & - \,
\mu^2 \left[ \frac{d^2 h}{d \tau^2} + \frac{2}{\tau} \frac{dh}{d
\tau} \right] = -  \, \frac{\mu^2}{\tau^2} \, \frac{d \;}{d \tau}
\left( \tau^2  \frac{d h}{d \tau} \right) \, , \; \alpha =
\overline{1,3} \; , \nonumber
\end{eqnarray}
with no summation on the repeated indices, and plugging the (82)
in, it yields
\begin{eqnarray} R_{\alpha \beta \alpha \beta} & = &
- \, \mu^2 \left[ G_0 + \kappa_0 \int \frac{d \tau}{\tau } \left(
\frac{d \psi_+}{d \tau} \right)^2 \right] \nonumber \\* R_{\alpha
4 \alpha 4} & = & - \, R_{\alpha \beta \alpha \beta } \, + \,
\frac{\kappa_0}{2} \, \mu^2 \, \left( \frac{d \psi_+}{d \tau}
\right)^2
\end{eqnarray}
On the other hand, completely independent of the form of $h$, the
same curvature components can be derived from the exact form (26)
of the Einstein-(generalized) Gordon equations, where the
potential (13) reads, within the linear approximation assumption,
\[
\left. \left. V( \chi ) = V( \chi_M + \psi ) = V( \chi_M ) +
\frac{dV}{d \chi } \right|_{\chi_M} \psi + \frac{1}{2}
\frac{d^2V}{d \chi^2} \right|_{\chi_M} \psi^2 + {\cal O}\left(
\psi^{n > 2} \right)
\]
i.e., since $\chi_M$ is an extremum of $V$,
\begin{equation}
V( \psi ) \, = \, \frac{3 \mu^4}{2 \lambda} \, - \,
\frac{\mu^2}{2} \, \psi^2
\end{equation}
Thus, one gets the relations
\begin{eqnarray}
\left( f_{|4} \right)^2 - \, \frac{e^{-2f}}{a^2} & = &
\frac{\kappa_0}{6} \left[ \left( \psi_{|4} \right)^2 \, - \, \mu^2
\, \psi^2 \right] \nonumber \\* 2 \left[ f_{|44} + \left( f_{|4}
\right)^2 \right] & = & - \, \frac{\kappa_0}{2} \left[ \left(
\psi_{|4} \right)^2 \, + \, \mu^2 \, \psi^2 \right] - \left[
\left( f_{|4} \right)^2 - \, \frac{e^{-2f}}{a^2} \right]
\end{eqnarray}
which straightforwardly lead, because of (21, with $k=-1$), to the
entirely $\psi_+ - $dependent expressions of the essential curvature
components,
\begin{eqnarray}
R_{\alpha \beta \alpha \beta} & = & \frac{\kappa_0}{6} \left[
\left( \frac{d \psi_+}{dt} \right)^2 - \, \mu^2 \, \psi_+^2
\right] \nonumber \\* R_{\alpha 4 \alpha 4 } & = &
\frac{\kappa_0}{6} \left[ 2 \left( \frac{d \psi_+}{dt} \right)^2 +
\, \mu^2 \, \psi_+^2 \right] \nonumber
\end{eqnarray}
(no summation, $ \alpha \neq \beta \in \lbrace 1,2,3 \rbrace$),
which can obviously be written, in terms of the dimensionless
variable $\tau$, as
\begin{eqnarray}
& (a) & R_{\alpha \beta \alpha \beta} \; = \; \frac{\kappa_0}{6}
\, \mu^2 \left[ \left( \frac{d \psi_+}{d \tau } \right)^2 - \,
\psi_+^2 \right] \nonumber \\* & (b) & R_{\alpha 4 \alpha 4 } \; =
\; \frac{\kappa_0}{6} \, \mu^2 \left[ 2 \left( \frac{d \psi_+}{d
\tau } \right)^2 + \, \psi_+^2 \right]
\end{eqnarray}
Adding the two expressions, we instantly get the second result
(83) and that is a very good cheek-out since, basically, the
formulae (86) and (83) have been independently derived, speaking
of the concretely employed methods. Hence, it is quite sufficient
to refer the calculations just to the spatial sectional curvature
components and, if we worked well, the two expressions (86.a) and
(the first of) (83) should produce the same result. As, because of
(75), we would be dealing with the three Bessel functions, $\left
\lbrace I_n (\tau ) \right \rbrace_{n= \overline{0,2}}$, the
analytical closed-form estimation of the integral involved in (83)
would certainly be out of (the normal) reach. So that, we are
going to consider only the first two terms, i.e.
\[
I_1 ( \tau ) \cong \frac{\tau}{2} \, + \, \frac{\tau^3}{16} \, ,
\]
from the power-series expression of the modified Bessel function
$I_1$. Consequently,
\begin{equation}
\psi_+ \, = \, \frac{{\cal N}_+}{2} \left[ 1 + \frac{\tau^2}{8}
\right] , \; \frac{d \psi_+}{d \tau} \, = \, \frac{{\cal N}_+}{8}
\, \tau
\end{equation}
and therefore, the first of (83) becomes
\[
R_{\alpha \beta \alpha \beta } \, = \, - \, \mu^2 \, G_0 \, - \,
\frac{\kappa_0 \mu^2}{128} \, {\cal N}_+^2 \, \tau^2 \, ,
\]
while (86.a) does concretely read
\begin{equation}
R_{\alpha \beta \alpha \beta } \, = \, - \, \frac{\kappa_0
\mu^2}{24} \, {\cal N}_+^2 \, - \, \frac{\kappa_0 \mu^2}{128} \,
{\cal N}_+^2 \, \tau^2
\end{equation}
Hence, as we have said, the integration constant $G_0$, in (82),
is just seemingly arbitrary for it must actually equate the static
contribution $\frac{\kappa_0}{24}{\cal N}_+^2$ of the perturbation
field $\psi_+$. Subsequently, either from (83) or straightly from
(86.b), the mixed components of the perturbed curvature, close to
the singular epoch $t=0$, are given by
\begin{equation}
R_{\alpha 4 \alpha 4} \, = \, \frac{\kappa_0 \mu^2}{24} \, {\cal
N}_+^2 \, + \, \frac{\kappa_0 \mu^2}{64} \, {\cal N}_+^2 \, \tau^2
\end{equation}
and, just for the sake of completeness, the metric perturbation
function reads
\begin{equation}
h \, = \, - \, \frac{\kappa_0 {\cal N}_+^2}{144} \left[ 1 \, + \,
\frac{9 \tau^2}{80} \right] \, \tau^2
\end{equation}
It can be concluded so far, inspecting the ^^ ^^ early''
coherently perturbed components (88), (89) of the curvature
tensor, that the spontaneously inner-parity breaking generated
Milne phase is clearly unstable, no matter how small the
source-field perturbation is, and it primarily runs into an
anti-de Sitter phase of scalar curvature $R[0] = - \,
\frac{\kappa_0}{2} \left( \mu {\cal N}_+ \right)^2$.

\section{Higgs$-$anti-de Sitter Spacetime Bubbles}

That is nicely closing the circle, for it brings us back to the
only Einstein equation (34.a) characterizing the spacetime
^^ ^^ supported'' by the other two fixed point values of the field
$\chi$. Written with respect to the cosmological scale function
(39) and introducing the notation
\begin{equation}
\omega_0^2 \, = \, \frac{\kappa_0 \mu^4}{2 \lambda} \; \;
\Leftrightarrow \; \; \omega_0 = \mu^2 \sqrt{ \frac{ \kappa_0}{2
\lambda}} \, ,
\end{equation}
the above equation, (34.a), becomes extremely simple
\begin{equation}
\left( \frac{dS}{dt} \right)^2 = \, 1 - \, \omega^2_0 \,
S^2 \;
\; \Rightarrow \; \; \frac{dS}{dt} \, = \, \pm \,
\sqrt{1-(
\omega_0 S)^2 } \, ,
\end{equation}
so that, its general solution reads, ^^ ^^ by the book'',
\begin{equation}
S(t) \, = \, \omega_0^{-1} \, \sin ( \omega_0 t + \gamma_0 ) \, ,
\end{equation}
where the constant phase-factor $\gamma_0$ accounts for both the
sign-choices ($\pm$). Actually, considering a positive scale
factor ^^ ^^ $a$'' $-$ with physical dimension of length $-$
and because $f : {\rm{\bf R}} \to {\rm{\bf R}}$, the scale
function defined by (39) must be non-negative, reading therefore
\begin{equation}
S(t) \, = \, \frac{1}{\omega_0} | \sin (\omega_0 t + \gamma_0 ) |
, \; \; {\rm such \; that} \; \; f = \ln | \sin (\omega_0 t +
\gamma_0 ) |
\end{equation}
Hence, for the other two fixed-point values $\chi_{L,R}$ given by
(32), double roots of the potential (31), we have been through
quite fast with the non-linear Einstein-Gordon system (30), once
we had (34.a) integrated, getting its exact solution(s) as a pair
of anti-de Sitter Universes, whose metric does explicitly read (in
terms of $(k=-1)-$spherical coordinates $\lbrace r, \theta) ,
\varphi \rbrace$
\begin{equation}
ds^2 \, = \, \frac{1}{\omega_0^2} \, \sin^2 ( \omega_0 t +
\gamma_0 ) \left[ \frac{(dr)^2}{1+r^2} + r^2 \, d \Omega^2
\right] - \, (dt)^2 \, ,
\end{equation}
actually representing two harmonically oscillating
$(k=-1)-$bubbles, that go through an eternal sequence of cosmic
Bangs and Crunches, one of them filled up with the remnant field
$\chi_L = - 2 \mu \sqrt{6/ \lambda }$ and the other seemingly
empty as the massive source-field $\chi$ vanishes everywhere
inside, but not before it left an enormous amount of exotic
vacuum-energy. In this respect, let us see what the numbers would
be if one took $\lambda \approx 6$ and considered the smallest
symmetry breaking scale, namely the one involved in the Higgs
sector of the Standard Model, where (probably, for now, as the
Higgs has not been experimentally detected yet) its mass, $m_H =
\sqrt{2} \mu$, lies somewhere inbetween 115 and 300 $GeV$,
i.e. (in Kilos)
\begin{equation}
2 \cdot 10^{-25} \; (kg) \leq m_H = \sqrt{2} \, \mu < 5.(3) \cdot
10^{-25} \; (kg)
\end{equation}
First, with the fundamental (universal) constants $c$ and
$\hbar$ plugged in, the vacuum-energy density
\begin{equation}
{\cal H}_0 \, = \, T_{44} [0] = \, - \, \frac{3 \mu^4}{2 \lambda}
\, ,
\end{equation}
sustaining the anti-de Sitter ^^ ^^ bubble'' where the Higgs cools
down to its undynamized ground state $\chi = 0$, does explicitly
read
\begin{equation}
{\cal H}_0 \, = \, - \, \frac{3 m_H^4 c^5}{8 \lambda \hbar^3} \,
= \, - \, \frac{1.246  \div 63}{\lambda} \cdot 10^{45} \;
({J/m^3}) ,
\end{equation}
yielding in modulus, for $\lambda \approx 6$, the impressive
values
\begin{equation}
| {\cal H}_0 | \cong 2 \cdot 10^{44} \div 10^{46} \; (J/m^3)
\end{equation}
which, nevertheless, have been frequently encountered in the
Domain Walls Theory. Similarly, the proper pulsation (91),
measured in $s^{-1}$, is given by the formula
\begin{equation}
\omega_0 \, = \, \frac{c \, m_H^2}{\hbar} \sqrt{\frac{\pi
G c}{\hbar \lambda}} \, \cong \, \frac{2.94 \div
20.82}{\sqrt{\lambda}} \cdot 10^9 \; (s^{-1})
\end{equation}
that subsequently leads to the proper frequency
\begin{equation}
\nu_0 \, = \, \frac{c \, m_H^2}{2 \pi \hbar} \sqrt{\frac{\pi G c}{
\hbar \lambda}} \cong \frac{0.468 \div 3.314}{\sqrt{\lambda }}
\; \; (GHz)
\end{equation}
and to the geometrical period of the Bang-Crunch cycles in these
Higgs-anti-de Sitter spacetime bubbles,
\begin{equation}
T \, = \, \frac{\pi}{\omega_0} \, \cong \, ( 0.151 \div 1.071 )
\sqrt{\lambda} \; (ns) \; .
\end{equation}
For the considered $\lambda$, their respective values are
\begin{eqnarray}
& (a) & \omega_0 \cong (1.2 \div 8.5 ) \cdot 10^9 \; (s^{-1}) \, ,
\nonumber \\* & (b) & \nu_0 \cong 0.19 \div 1.35 \; (GHz) \, ,
\nonumber \\* & (c) & T \cong 0.35 \div 2.626 \; (ns) \, .
\end{eqnarray}

Concerning the cosmological length-scale parameter
$\omega_0^{-1}$, which is nothing else but the amplitude of the
anti-de Sitter scale function oscillation, it reads (in
international units)
\begin{equation}
\omega_0^{-1} \, = \, \frac{\hbar}{m_H^2} \sqrt{\frac{\hbar
\lambda}{\pi Gc}} \cong \frac{1.44 \div 10.2}{\sqrt{\lambda }} \;
\; (cm)
\end{equation}
and is getting, as $\omega_0$ did in (103.a), the numerical values
\begin{equation}
\omega_0^{-1} \cong 3.53 \div 25 \; (cm)
\end{equation}
Based on these data, we can speculate a bit, in a sort of {\it
what if...}-manner, on the possible cosmological consequences of
the existence, in some regions of our Universe, of some
(2+1)-dimensional windows towards the bulk-space extra-dimensions
where such Higgs--anti-de Sitter (harmonically oscillating)
bubbles might be living. For instance, the upper limit of the
proper frequency $\nu_0$ is pretty close to the famous 21 $cm(s)$
Hydrogen-line so that, inspecting the whole sky, if the
Higgs-boson mass were around 300 $GeV$, there would (presumably)
be some conventionally unexplained deviations from the averaged
level of the electromagnetic radiation received from the known and
ordinary excited baryonic astrophysical objects. Similarly,
referring to the rest of the $\nu_0-$values, as the present
thermalized-photons temperature is too small to significantly
dynamize the Higgs-like field $\chi$ around its ground state,
$\chi =0$, one can presume that, watching for instance the Giant
Voids, which are pretty much deprived of the other forms of
conventional matter, there might be detected some disturbances in
(or, eventually fluctuating, anisotropy of) the $\lbrace n \,
\nu_0 \rbrace_{n= \overline{1,3}}$ channels of the Cosmic
Microwave Background Radiation, coming from the junction with
(such) an electroweak spontaneously broken$-$anti-de Sitter
^^ ^^ small'' scale Universe. In some respect, the situation is very
much alike the one in Chaotic Inflation $-$ where inflating Baby
Universes pop up (chaotically) from the spacetime foam $-$ except
that now we deal with harmonically oscillating $(k=-1)-$regions,
geometrodynamically exactly accommodating the initial
self-interacting field $\Phi$ in one of the absolute minima of its
quartic potential, that pop up in an already inflated bubble,
which is our own Universe. The reason why we have included the
third harmonic of $\nu_0$ among the frequencies on which there
might be some deviations from the black body radiation law of the
Cosmic Microwave Background, lies in the manner the total energy
of an anti-de Sitter three-dimensional ball depends upon time.
Indeed, considering the well-known formula (for the energy of a
$\lbrace t = cst. \rbrace-$compact filled in by the matter-density
${\cal H}_0$ )
\begin{equation}
E = \int_{N_3 ( t=cst.)} \sqrt{-g} \; {\cal H}_0 \; d^3 x \, ,
\end{equation}
where $d^3 x = dr \, d \theta \, d \varphi$, $\sqrt{-g} = \frac{|
\sin^3 (\omega_0 t)|}{\omega_0^3} \, \frac{r^2 \sin
\theta}{\sqrt{1+r^2}}$ (derived from (95), discarding $\gamma_0$)
and ${\cal H}_0$ being given by (98), it yields for the
instantaneous (total) energy of the $\lbrace t=cst.
\rbrace -H^3-$ball, of dimensionless radius $r_0$, the expression
\begin{equation}
E(t) \, = \, - \, {\cal E} F(r_0) \, | \sin (\omega_0 t )|^3 \, ,
\end{equation}
where the radial volume-function $F(r_0)$ and the (physically
dimensional) energy amplitude ${\cal E}$ are respectively given by
\begin{eqnarray}
& (a) & F(r_0) \, = \, = \, \frac{1}{2} \left[ r_0 \sqrt{1+r_0^2}
\, - \, \ln \left( r_0 + \sqrt{1+r_0^2} \right) \right] ,
\nonumber
\\* & (b) & {\cal E} \, = \, \frac{3 \hbar c^4}{2 G m_H^2}
\sqrt{\frac{\hbar \lambda}{ \pi G c}} \, , \; [{\cal E}]= Joule
\end{eqnarray}
Since $\sin^3 x \equiv \frac{3}{4} \sin x - \frac{1}{4} \sin
(3x)$, it clearly results a $25 \% $ energy-distribution on the $3
\, \nu_0-$channel. With (108) and (107), the mean-energy during an
anti-de Sitter cycle (102),
\[
\langle E \rangle_T \, = \, \frac{1}{T} \int_0^T E(t) \, dt \, =
\, - \, \frac{{\cal E}F(r_0)}{\pi} \int_0^{\pi} \sin^3 \gamma \, d
\gamma \, ,
\]
i.e. $-$ formally $-$
\begin{equation}
\langle E \rangle_T = \, - \, \frac{4 {\cal E}}{3 \pi} \, F(r_0)
\, ,
\end{equation}
does actually read
\begin{equation}
\langle E \rangle_T = \, - \, \frac{2 \hbar c^4}{\pi G m_H^2} \,
\sqrt{\frac{\hbar \lambda}{\pi G c}} \, F(r_0) \, ,
\end{equation}
and, in terms of absolute values, it already gives an idea about
the effectively involved power
\[
{\cal P}_{ef} \, = \, \frac{| \langle E \rangle_T|}{T} \, ,
\]
namely, in $watts$,
\begin{equation}
{\cal P}_{ef} \, = \, \frac{2c^5}{\pi^2 G} \, F(r_0)
\end{equation}
Of course, in a rigorous manner, the instantaneous power ${\cal
P}(t)$ comes being expressed from (107) as
\[
{\cal P} \, = \, \frac{dE}{dt} \, = \, - \, 3 \, \omega_0 \,
{\cal E}
F(r_0) \sin^2 ( \omega_0 t) \cos ( \omega_0 t) \, , \; ( \forall )
\; t \in \left[ 0, \, \frac{\pi}{\omega_0} \right] ,
\]
i.e. $-$ inserting (100) and (108.b) $-$
\begin{equation}
{\cal P} \, = \, - \, \frac{3c^5}{2G} \, \sin^2 (\omega_0 t) \cos
(\omega_0 t) \, F(r_0) \, ,
\end{equation}
so that it takes symmetric values during an anti-de Sitter cycle,
being negative in its first half, when the bubble blows to its
maximum size  $\omega_0^{-1} r_0$, and respectively positive, on
the second half, while the deflating bubble goes into the
$T$-crunch. Hence, although the sooth averaged power is zero, yet
one can meaningfully define the {\it anti-de Sitter semi cycle
mean-power}, (in absolute value),
\begin{equation}
\langle {\cal P} \rangle_{1/2} \, = \,
\frac{3 c^5}{2 G} \left[ \frac{2 \omega_0}{\pi}
\int_{0}^{\frac{\pi}{2 \omega_0}} \sin^2 (\omega_0 t) \cos( \omega_0 t)
d t \right] \, F(r_0) \, \, ,
\end{equation}
i.e.
\begin{equation}
\langle {\cal P} \rangle_{1/2} \, = \,
\frac{c^5}{\pi G} \, F(r_0) \; (W) \, ,
\end{equation}
which is released for instance in the crunch-directed decaying phase;
compared to ${\cal P}_{ef}$, it is $\frac{\pi}{2}$-times larger but,
nevertheless, of the same order of magnitude.

At the electroweak symmetry breaking scale, that has been
considered here, the {\it energy amplitude} ^^ ^^ alone'' already
achieves {\it intriguing} numerical values,
\begin{equation}
{\cal E} \, \cong \, (0.7 \div 4.8) \cdot 10^{43} \; (J) \, ,
\end{equation}
which are $-$ ^^ ^^ astrophysically speaking'' $-$ of the same
order (of magnitude) with the ones of a {\it medium size} galaxy.
Hence, speculating again, envisaging the modulus of the
mean-energy (110),
\begin{equation}
| \langle E \rangle_T | \, \cong \, (0.3 \div 2) F(r_0) \cdot
10^{43} \; (J) \, ,
\end{equation}
it might turn out that decaying Higgs-vacuum$-$anti-de Sitter bubbles,
no larger
than few tens of $\omega_{0}^{-1}$ (given by (105)), could (in
principle)
provide enough energy to act as {\it seeds} in the galaxy formation
process.
In what it concerns the power (114),
\begin{equation}
\langle {\cal P} \rangle_{1/2} \, \cong \, F(r_0) \cdot 10^{52} \; (W)
\, ,
\end{equation}
which would be released if the bubble stopped growing again after
it crunched, that might account for the one emitted by quasars, if
some understandable anti-de Sitter-Higgs$-$electromagnetic
conversion (mechanism), acting in the core of the quasar, could be
figured out. Nevertheless, it should exist, since the derived
power expressions (111-114) are completely independent not only of
the electroweak breaking scale parameters, but also of the
universal Planck constant, being therefore entirely of generally
relativistic gravitational origin.

\section{${\mbox{\boldmath $S^2$-}}$Cobordism and ^^ ^^ Wick Companions''}

Finally, there are two more features (of the topic we are
discussing) that we would like to address in the remaining part of
the paper.

The first concerns the $\lbrace r = cst. \rbrace -(2 \, + \,
1)$-dimensional cobordism of the anti-de Sitter sphere of
coordinate-radius $r_0$ to a spatially flat FRW-Universe of scale
function $a(T)$. It comes about by equating the corresponding
metrics,
\begin{equation}
\frac{r_{0}^{2}}{\omega_{0}^{2}} \, \sin^2 (\omega_0 t) \,
d \Omega^2 \, - \, (d t)^2 \, = \, a^2 (T) (d R)^2 \, +
\, a^2 (T) R^2 d \Omega^2 \, - \, (d T)^2 \, \, ,
\end{equation}
such that the {\it first cobordering equation} reads
\begin{equation}
a(T) R \, = \, \frac{r_0}{\omega_0} \, | \sin(\omega_0 t) | \, \, ,
\end{equation}
leading therefore, to the {\it second} one
\[
(d T)^2 \, - \, a^2 (T) (d R)^2 \, = \, (d t)^2 \, \, ,
\]
i.e.
\begin{equation}
\left( \frac{d T}{d t} \right)^2 \, - \,
a^2 (T) \left( \frac{d R}{d t} \right)^2 \, = \, 1 \, \, .
\end{equation}
Extracting $R$ from (119) and taking its derivative with respect
to $t$, then plugging the result back into (120), the latter
becomes
\begin{equation}
\left[ 1 - \frac{r_{0}^{2} \sin^2 (\omega_0 t)}{\omega_{0}^{2} \,
a^2(T)}
\left( \frac{d a}{d t} \right)^2 \right]
\left( \frac{d T}{d t} \right)^2 +
\frac{r_{0}^{2} \sin(2 \omega_0 t)}{\omega_0 \, a(T)} \frac{d a}{d t}
\frac{d T}{d t} \, - \left[ 1 + r_{0}^{2}
\cos^2 (\omega_0 t) \right] = 0
\end{equation}
and, as it can be noticed, although is just a first-order
differential equation, it actually is a highly nonlinear one,
especially when general forms of the $(k=0)-$scale function $a(T)$
are to be considered. Moreover, because of the trigonometric
functions involved in each of the three terms, (the dimensionless
coordinate-radius) $r_0$ {\it alone} is getting us in trouble,
even for simple forms of $a(T)$, particularly when it achieves
{\it large} values. Hence, a {\it closed form exact solution} to
the second cobordering equation (121) does not come easy.

However, some particular $-$ but not trivial $-$ cases can be
worked out completely, even if they might be looking a bit nasty,
and we are talking here about the ^^ ^^ critical'' case where
\begin{equation}
\frac{1}{a} \frac{d a}{d t} \, = \,
\frac{\omega_0}{r_0} \, | \sin ( \omega_0 t) |^{-1} \, \, ,
\end{equation}
such that it reduces the degree of (121), regarded as an algebraic
equation in $\left( \frac{d T}{d t} \right)$, yielding the far
much simpler equation
\begin{equation}
\frac{d T}{d t} \, = \, \frac{1 + r_{0}^{2} \cos^2 (\omega_0 t)}{2
r_0 \cos (\omega_0 t)} \, \, ,
\end{equation}
whose solution
\begin{equation}
T \, = \, T_0 \, + \, \omega_{0}^{-1} \left[ \frac{1}{2 r_0}
\ln \left| \frac{1 + {\rm tg} \left(
\frac{\omega_0 t}{2} \right)}{1 - {\rm tg}
\left( \frac{\omega_0 t}{2} \right)}
\right| \, + \, \frac{r_0}{2} \sin ( \omega_0 t) \right] \, , \, \,
T_0 = cst. \in {\rm{\bf R}} \, \, ,
\end{equation}
gives the concrete dependence (in this case) of the $(k = 0)-$RW
universal time on the one in the anti-de Sitter bubble.
Because of (123), the $(k = 0)-$scale function equation (122) reads
\[
\frac{1}{a} \, d a \, = \, \left( \frac{1}{r_{0}^{2}} +
\frac{1}{2} \right) \frac{d( \omega_0 t)}{\sin (2 \omega_0 t)} +
\frac{1}{2} \, {\rm ctg} (2 \omega_0 t) \, d( \omega_0 t) \, \, ,
\]
getting therefore the solution
\begin{equation}
a(t) \, = \, a_0 [1 - \cos ( 2 \omega_0 t)]^{\frac{1}{4}} \,
| {\rm tg} ( \omega_0 t) |^{- \frac{1}{2 r_0^2}} \, , \, \,
a_0 = cst. \in {\rm{\bf R}}_{+} \, \, ,
\end{equation}
which, together with (124), give the complete parametric
representation of the scale function $a(T)$. Consequently, the
first cobordering equation (119) does explicitly set the behaviour
of the ^^ ^^ true'' radius $R$ of the anti-de Sitter sphere as it
is actually seen from (within) the corresponding spatially flat
Universe (of scale function derived from) (125) and (124); that is
\begin{equation}
R(t) \, = \, \frac{\omega_{0}^{-1} r_0}{2^{1/4} a_0}
| \sin (\omega_0 t) |^{\frac{1}{2}} \,
| {\rm tg} ( \omega_0 t) |^{- \frac{1}{2 r_0^2}} \, \, .
\end{equation}
These are non-perturbative (exact) results. Considering now a
small enough anti-de Sitter bubble, such that $r_0^2 << 1$, they
got major simplification since (124) becomes
\begin{equation}
2 r_0 \, \omega_0 (T - T_0) \cong
\ln \left| \frac{1 + {\rm tg} ( \frac{\omega_0 t}{2})}
{1 - {\rm tg} ( \frac{\omega_0 t}{2})} \right| \, ,
\end{equation}
so that
\[
{\rm tg} (\frac{\omega_0 t}{2}) \, = \, {\rm th}
[ r_0 \, \omega_0 (T - T_0)] \, ,
\]
which leads, after a short calculation, to the scale function
expression
\begin{equation}
a(T) \, = \, 2^{1/4} a_0 \, | {\rm th}
[2 r_0 \omega_0 (T - T_0)] |^{\frac{1}{2}} \,
| {\rm sh} [2 r_0 \omega_0 (T - T_0)] |^{\frac{1}{2 r_0^2}}
\end{equation}
and to the $(k = 0)-$radial coordinate evolution law (in RW-time)
\begin{equation}
R(T) \, = \frac{\omega_0^{-1} r_0}{2^{1/4} a_0} \,
| {\rm th} [2 r_0 \, \omega_0 (T - T_0)] |^{\frac{1}{2}} \,
| {\rm sh} [2 r_0 \, \omega_0 (T - T_0)] |^{- \frac{1}{2 r_0^2}}
\end{equation}
Late into the future, for $T - T_0 >> (2 \, r_0 \,
\omega_0)^{-1}$, each of them does respectively go as
\begin{eqnarray}
&(a)& \, a(T) \, \cong \, b_0 \, \, e^{\frac{\omega_0}{r_0} T} \,
\, , \nonumber \\* &(b)& \, R(T) \, \cong \, \frac{\omega_{0}^{-1}
r_0}{b_0} \, \, e^{- \frac{\omega_0}{r_0} T} \, \, ,
\end{eqnarray}
with $b_0 = a_0/2^{\frac{1}{2 r_0^2}}$, lighting up clearly a strongly
decaying Higgs-vacuum (small scale) bubble, $S^{2}-$connected to an
extremely
fast inflating universe.

As a matter of fact, this beautiful picture can also be obtained
as an ^^ ^^ uncritical'' {\it exact solution} of the cobordering
equation (121) in the case where the physical radius
$\omega_0^{-1} r_0$ of the anti-de Sitter sphere does sharply
equate the inverse, $H_0^{-1}$, of the Hubble constant of a de
Sitter Steady-State Universe,
\begin{equation}
a(T) \, = \, e^{H_0 T} \; , \, \, H_0 > 0 \, \, .
\end{equation}
So, using (131) and the aforementioned coordinate-radius constraint,
\begin{equation}
r_0 \, = \, \frac{\omega_0}{H_0} \, \, ,
\end{equation}
the equation (121) gets the much more tractable form
\begin{equation}
\left[ \cos \tau \, \frac{d {\cal T}}{d \tau} \, + \,
\frac{\omega_0}{H_0} \sin \tau \right]^2 - \left( 1 +
\frac{\omega_0^2}{H_0^2} \right) = 0 \, ; \, \, \tau = \omega_0 t
\, , \, \, {\cal T} = \omega_0 T \, ;
\end{equation}
whose {\it time-orientation preserving} positive branch,
\begin{equation}
\cos \tau \, \frac{d {\cal T}}{d \tau} \, + \,
\frac{\omega_0}{H_0} \sin \tau \, = \, \sqrt{1 + \left(
\frac{\omega_0}{H_0} \right)^2} \, \, ,
\end{equation}
does immediately lead to the de Sitter$-$anti-de Sitter
{\it synchronization} law
\[
{\cal T} - {\cal T}_0 = \frac{\omega_0}{H_0}
\ln | \cos \tau | + \sqrt{1 + \frac{\omega_0^2}{H_0^2}}
\ln \left| \frac{1 + {\rm tg} (\tau/2)}{1 - {\rm tg} (\tau/2)} \right|
\, \, ,
\]
i.e., in physically dimensional quantities,
\begin{equation}
T - T_0 = H_0^{-1} \left[
\ln | \cos (\omega_0 t) | +
\sqrt{ \left( \frac{H_0}{\omega_0} \right)^2 + 1} \,
\ln \left| \frac{1 + {\rm tg} \left( \frac{\omega_0 t}{2} \right)}
{1 - {\rm tg} \left( \frac{\omega_0 t}{2} \right)} \right| \right]
\end{equation}
and subsequently, through (131) and the first cobordering equation
(119), to the variation law of the de Sitter-radial coordinate,
\begin{equation}
R(t) \, = \, H_0^{-1} | {\rm tg} (\omega_0 t) | \cdot
\left[ \frac{| \cos (\omega_0 t) |}{1 + \sin (\omega_0 t)}
\right]^{\sqrt{1 + (H_0/ \omega_0)^2}} \, \, .
\end{equation}
As it can be noticed, using (135) written as
\begin{equation}
\frac{ | \cos ( \omega_0 t) |^{\sqrt{1 + (H_0/ \omega_0)^2} - 1}}
{\left[ 1 + \sin ( \omega_0 t) \right]^
{\sqrt{ 1 + (H_0/ \omega_0)^2}}} \, = \, e^{- H_0 T} \, \, ,
\end{equation}
where we have set (for convenience) $T_0 = 0$, the expression (136)
reads
\begin{equation}
R \, = \, H_0^{-1} | \sin (\omega_0 t) | e^{- H_0 T}
\end{equation}
and asymptotically achieves the (130.b)-like behaviour,
\begin{equation}
R(T) \cong H_0^{-1} \, e^{- H_0 T} \, \, ,
\end{equation}
at late events (into the future), where, as $T \to \infty$,
$| \sin ( \omega_0 t) | \to 1$.

However, in the general case, where $r_0^2 << 1$ is clearly
invalidated, we could not find other exact solutions, in {\it
closed-form}, besides the ones given above; eventually, a
numerical study of the cobordering equations (121), (119) for
power-like scale functions, $a \sim T^{\nu}$, with $\nu > 0$, such
as the ones in the radiation dominated era, $\nu = 1/2$, or in the
one of ^^ ^^ dusty''-matter, $\nu = 2/3$, even supplied with an
accelerating cosmological term, might be quite important for it
could point out some sort of bifurcations in the $(k =
0)-$cosmological evolution of {\it large} Higgs$-$ anti-de Sitter
spactime bubbles. What else could be done in this respect, would
be to look for the proper {\it simultaneous} embeddings of the two
$S^{2}-$connected universes, so that to get a clear and ^^ ^^ very
pictural'' understanding of the resulting spacetime {\it global}
structure; that can further be used as the {\it unperturbed}
background in similar $-$ but seriously improved $-$ models with
conventional matter-sources and {\it dynamical} ^^ ^^ remnant''
field.

The second matter we would have liked to refer to had regarded the
{\it instanton} companion (gotten by a Wick-rotation) of the
Higgs$-$anti-de Sitter spacetime, which is precisely the
^^ ^^ never-started $-$ never-ending'' $(k = 1)-$de Sitter Universe,
\begin{equation}
d s^2 = \frac{1}{\omega_0^2} \, {\rm ch}^{2}(\omega_0 t)\,
d l_{S^3}^{2} \, - \, (d t)^2 \,\, , \;
\omega_0 = \mu^2 \, \sqrt{\frac{\kappa_0}{2 \lambda}} \, ,
\end{equation}
in the (static source-field) {\it fixed point} case, and does
beautifully turn into the Linde's {\it Inflationary} Universe with
quadratic driven-source, $\mu^2 \, \phi^2$, where $\phi$ is a
genuine {\it inflaton}, when the spontaneous $Z_2-$invariance
breaking resulting field gets excited.

\begin{flushleft}
\begin{Large}
{\bf Acknowledgement}
\end{Large}
\end{flushleft}
The warm hospitality, enjoyable atmosphere and fertile environment
of the Institute of Theoretical Science, University of Oregon are
deeply acknowledged. Special thanks go to J. Isenberg and X. Tata
for useful discussions and fruitful suggestions.

\end{document}